\documentclass[12pt,preprint]{aastex}
\def\ssp{\def\baselinestretch{1.1}\small\normalsize}
\def\tighttable{\def\baselinestretch{1.1}}
\def\arcsec{''}
\def\asec{\ifmmode ''\!. \else $''\!.$\fi}
\def\arcsecpoint{\ifmmode ''\!. \else $''\!.$\fi}
\def\micron{\ifmmode \mu{\rm m} \else $\mu$m\fi}

\def\Msun{M$_{\odot}$}

\def\HST{{\it HST}}
\def\HFF{{\it HFF}}
\def\sst{{\it Spitzer}} 
\def\chan{{\it Chandra}} 
\def\I{{F814W}}
\def\Y{{F105W}}

\def\J{{F125W}}
\def\JH{{F140W}}
\def\H{{F160W}}

\shorttitle{LBGs in MACS 0416}
\shortauthors{Infante et al.}
\slugcomment{Version 2.1, \today}
\received{2014 November 23}
\accepted{2015 **}
\articleid{}{} 

\begin{document}

\title{Young Galaxy Candidates in the Hubble Frontier Fields.\\II. MACS\,J0416$-$2403}

\author{
Leopoldo Infante\altaffilmark{1},
Wei Zheng\altaffilmark{2},
Nicolas Laporte\altaffilmark{1},
Paulina Troncoso\altaffilmark{1},
Alberto Molino\altaffilmark{12,14},
Jose M. Diego\altaffilmark{13},
Franz E. Bauer\altaffilmark{1,3,4},
Adi Zitrin\altaffilmark{6,7},
John Moustakas\altaffilmark{5},
Xingxing Huang\altaffilmark{8},
Xinwen Shu\altaffilmark{9},
David Bina\altaffilmark{15}, 
Gabriel B. Brammer\altaffilmark{16},
Tom Broadhurst\altaffilmark{10,11},
Holland C. Ford\altaffilmark{2},
Stefano Garc\'{i}a\altaffilmark{1},
Sam Kim\altaffilmark{1}
}
\altaffiltext{1}{Instituto de Astrof\'{\i}sica and Centro de Astroingenier\'{i}a, Facultad de F\'{i}sica,  Pontificia Universidad Cat\'{o}lica de Chile, Vicu\~{n}a Mackenna 4860, 7820436 Macul, Santiago, Chile}
\altaffiltext{2}{Department of Physics and Astronomy, Johns Hopkins University, Baltimore, MD 21218}
\altaffiltext{3}{Millennium Institute of Astrophysics, Vicu\~{n}a Mackenna 4860, 7820436 Macul, Santiago, Chile}
\altaffiltext{4}{Space Science Institute, Boulder, CO 80301}
\altaffiltext{5}{Department of Physics and Astronomy, Siena College, Loudonville, NY 12211}
\altaffiltext{6}{Cahill Center for Astronomy and Astrophysics, California Institute of Technology, MC 249-17, Pasadena, CA 91125}
\altaffiltext{7}{Hubble Fellow}
\altaffiltext{8}{Department of Astronomy, University of Science and Technology 
of China, Hefei, Anhui 230026, China}
\altaffiltext{9}{Department of Physics, Anhui Normal University, Wuhu, Anhui,241000, China}
\altaffiltext{10}{Department of Theoretical Physics, University of Basque Country UPV/EHU, Bilbao, Spain}
\altaffiltext{11}{IKERBASQUE, Basque Foundation for Science, Bilbao, Spain}
\altaffiltext{12}{Instituto de Astronom\'ia, Geof\'isica e Ci\^encias Atmosf\'ericas, Universidade de S\~ao Paulo, Cidade Universit\'aria, 05508-090, S\~ao Paulo, Brazil}
\altaffiltext{13}{IFCA, Instituto de F\'isica de Cantabria, UC-CSIC, s/n. E-39005 Santander, Spain} 
\altaffiltext{14}{Instituto de Astrof\'isica de Andaluc\'ia - CSIC, Glorieta de la Astronom\'ia, s/n. E-18008, Granada, Spain}
\altaffiltext{15}{IRAP, CNRS - 14 Avenue Edouard Belin - F-31400 Toulouse, France}
\altaffiltext{16}{Space Telescope Science Institute, 3700 San Martin Drive, Baltimore, MD 21218, USA}
\setcounter{footnote}{16}

\begin{abstract}
We searched for $z\gtrsim 7$ Lyman-break galaxies (LBGs) in the
optical-to-mid-infrared Hubble Frontier Field and associated parallel
field observations of the strong-lensing cluster MACS\ J0416$-$2403. We
discovered 22 candidates, of which six lie at $z\gtrsim 9$ and one
lies at $z\gtrsim 10$. Based on the Hubble and Spitzer photometry, all
have secure photometric redshifts and a negligible probability of
being at lower redshifts, according to their peak probability ratios,
$\Re$. This substantial increase in the number of known high-redshift
galaxies allows a solid determination of the luminosity function at
$z\gtrsim 8$.  The number of high-z candidates in the parallel field
is considerably higher than that in the Abell 2744 parallel field.  
Our candidates have median stellar masses of $log(M_*) \sim
8.40^{+0.55}_{-0.31}$~\Msun, SFRs of $\sim 1.6^{+0.5}_{-0.4}
$~\Msun~yr$^{-1}$, and SFR-weighted ages of $\lesssim 310^{+70}_{-140}
$~Myr.  Finally, we are able to put strong constraints on the 
$z = 7,8,9$ and $10$ luminosity functions. 
One of the objects in the cluster field is a $z\simeq10$ candidate, 
with a magnification of $\mu\sim20\pm13$. 
This object is likely the faintest $z\sim10$ object known to date, allowing
a first look into the extreme faint-end ($L\sim 0.04L^*$) of the $z\sim10$
luminosity function.
\end{abstract}


\section{INTRODUCTION}
The Universe at $z\sim 10$ was approximately 330 $h^{-1}$ million
years old. At this time, galaxy assembly was well underway and the
intergalactic medium (IGM) was being reionized by UV radiation
(presumably from the first stars and compact objects).  These two
processes mark an important era in the evolution of the Universe. In
that sense, it is important to understand how different young
galaxies are from local ones, and how their numbers evolve with
redshift.

In the last few years the number of galaxies discovered at the highest
redshifts has increased dramatically \citep{bouwens, bouwens1, ellis,
  oesch13, garth, bouwens4, bradley}. For instance, in 2008 only a
handful of $z\gtrsim 7$ galaxies were known and none had been
discovered above $z \sim 8$; today these numbers have grown to
$\sim300$ for $z \sim 7$, $\sim$100 for $z \sim 8$ and even $\sim 10$
at $z \sim 10$.
Of particular interest is the latest result by the {\it Planck} collaboration 
on the optical depth to reionization, $\tau$. The CMB power spectrum is sensitive to $\tau$ both 
in the temperature power spectrum and in the polarization power spectrum. 
The CMB data is not sensitive to the particular model of reionization but is very sensitive 
to its average redshift. The latest results from {\it Planck} 
suggest a lower value for $\tau$ than previous estimates from {\it WMAP} data.
The new value for the optical depth ($\tau=0.066 \pm 0.016$) translates into
 a mean redshift of reionization of $z \approx 8$ \citep{planckcollXIII}. 
When combined with recent measurements of the Ly-$\alpha$ forest that suggest
that reionization of the inter-galactic medium was nearly complete
by $z \approx 6$ \citep{fan06}, a simple picture of the reionzation history can be put together in which 
reionization may have happened gradually between $z=10$ and $z=6$. 
The lack of galaxies beyond $z=10$ in our deep fields 
support this hypothesis; and we may be already proving the first galaxies 
emerging from the dark ages. Under the simple hypothesis that 
reionization happened gradually between $z=10$ and $z=6$ and that this 
reionization took place in a patchy form, we should expect a statistically 
significant variability in the number density of detected galaxies at $z \approx 10$ 
when comparing different deep fields (for instance from the 
Hubble Frontier Fields \cite[\HFF,][]{hff} program since some 
of them may reach further in redshift. 

The increase in the number of known high-redshift galaxy candidates
is mainly due to the advent of new instrumentation on the Hubble Space
Telescope (\HST) Wide-Field Camera~3/Infrared Channel
\citep[WFC3/IR,][]{wfc3}, coupled with constraints from the {\em
  Spitzer Space Telescope} Infrared Array Camera
\citep[IRAC,][]{irac}. How far can we push these instruments to probe
the first galaxies? At $z\gtrsim 10$, galaxies are extremely
faint. Most have intrinsic luminosities fainter than $M_{*} \simeq
-19.5$ and sizes of a few hundreds of parsecs \citep[see][]{zheng,
  bouwens2, coe, bradley, bouwens4, kawamata, zitrin14}.  To observe
them, one option is to take advantage of gravitational lensing by
massive galaxy clusters, which magnify and shear background galaxy
light as it passes through the cluster. To achieve detections, two
crucial elements are needed: high enough amplification, which is
provided by observing massive galaxy clusters, and accurate
photometric redshifts, which are obtained by observing a large number
of wavelength bands with accurate photometry. Of course, to retrieve
the intrinsic source properties of the distant objects requires very
good mass lens models of the cluster. This confluence of data have
been secured by two recent \HST\ cluster surveys, the Cluster Lensing
And Supernova survey with Hubble \citep[CLASH,][]{postman} and the
\HFF.

The \HFF\ campaign (560 \HST\ orbits of Director's Discretionary Time)
in particular was designed to take advantage of \HST's unsurpassed
spatial resolution and sensitivity. Four clusters of galaxies are being
observed in Cycles 21 and 22, and two more are expected in Cycle 23.
\footnote{See
  http://www.stsci.edu/hst/campaigns/frontier-fields/documents/HDFI\_SWGReport2012.pdf}
Observations are carried out with the \HST\ ACS camera in three
optical bands (F435W, F606W and F814W) and with \HST\ WFC3 in four NIR
bands (F105W, F125W, F140W and F160W), to limiting magnitudes of $\sim
29-30 AB$. These are (or will be) complemented by deep \sst\ and
\chan\ observations of these six clusters. The
\HST\ \HFF\ observations have now been completed for clusters Abell
2744 and MACS J0416$-$2403 (hereafter A2744 and M0416, respectively),
and have resulted in dozens of publications. Among these:
\cite{zheng14} reported the discovery of 24 Lyman-break candidates
between $7\lesssim z \lesssim 10.5 $ and a triple system with a
photometric redshift of $z_{\rm ph}\simeq 7.4$ in A2744;
\cite{zitrin14} reported the geometrical confirmation of a
multiply-lensed $z\sim 10$ object in A2744; \cite{laporte2}
communicated the discovery of four bright $z\sim8$ galaxies in M0416;
and \cite{McLeod} reported the discovery of $z\sim 8 - 9$ galaxies in
M0416 (see also, Atek et al., 2014, Ishigaki et al. 2014).

In this paper (the second in our discovery series), we report the
discovery of 22 $z\gtrsim 7$ galaxy candidates in the \HFF\ cluster
M0416 and parallel fields. We analyse their physical properties via
SED fitting and combine results from A2744 and MACS4016 to compute
luminosity functions (LFs) at redshifts $z \simeq 7 - 8 - 9 ~\& ~10$.  The
$z\sim8$ galaxy candidates in M0416 from \cite{laporte2} are also used
in our analysis.  In $\S$\ref{sec:data} we describe the observations
and the data; $\S$\ref{sec:selection} details how galaxies are
selected; the lens models and photometric redshift estimations are
presented in $\S$\ref{sec:lens}; and in $\S$\ref{sec:discussion} we
present an SED fitting analysis, compute LFs and
discuss our results.

For all calculations, we adopt a concordance cosmology with
$\Omega_M=0.3$, $\Omega_\Lambda=0.7$ and $h=H_0/100\,{\rm
  km\,s^{-1}\,Mpc^{-1}}=0.7$, and the AB magnitude system throughout.

\section{DATA}\label{sec:data}

\begin{figure}
   \centering
   \includegraphics[width=0.8\textwidth]{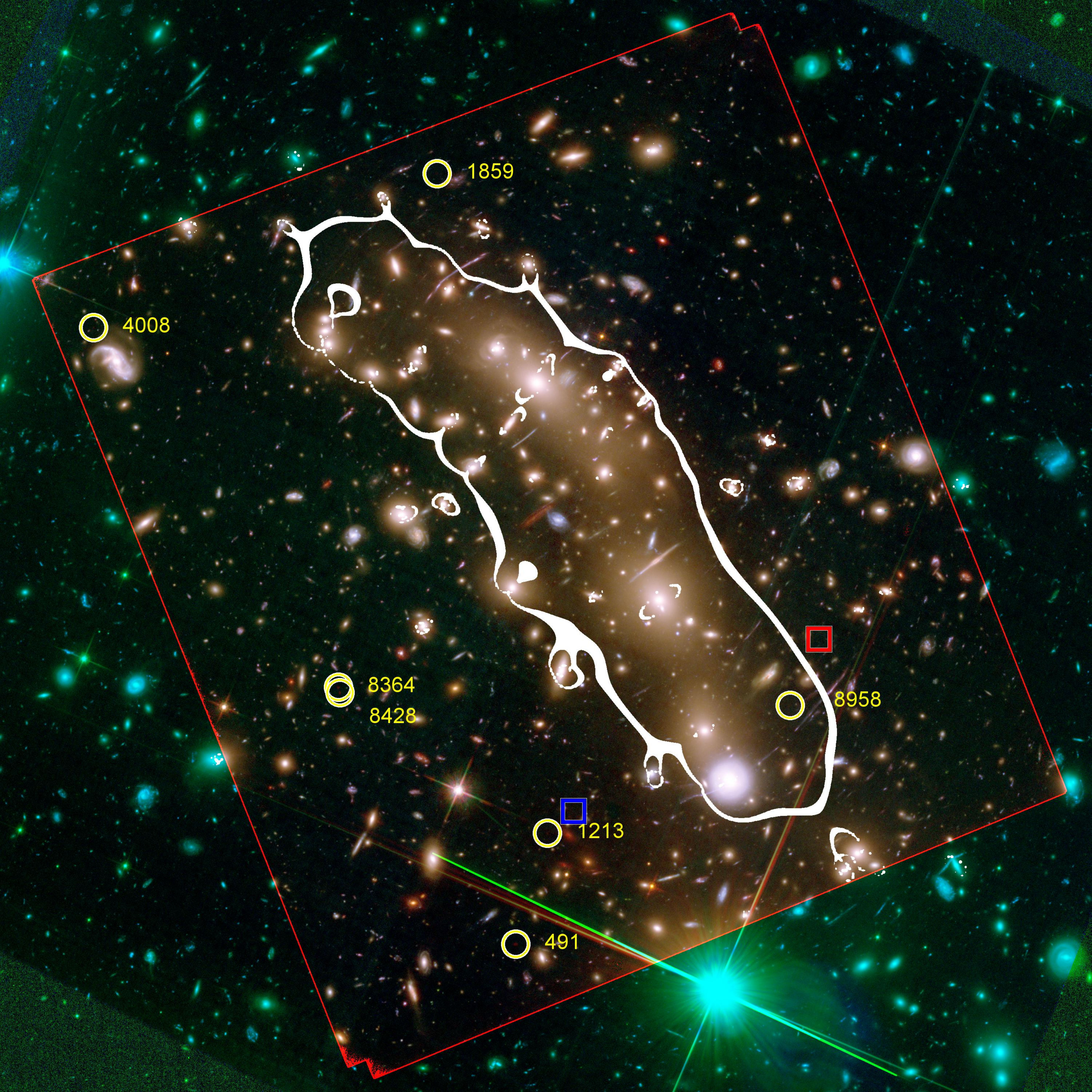}
\caption{A 3\arcmin$\times$3\arcmin \HST\ color image of the M0416
  cluster field, comprised of the ACS F606W (blue), ACS F814W (green),
  and WFC3 F160W (red) bands, with the $z=10$ critical curve from the
  \citet{diego15} lens mass model overlaid. Our $z>7$ candidates are
  labeled and marked with yellow circles. The red and blue squares are 
  model predictions for potential 8958 counter images. The typical error in the
  position for this solution is approximately 3 arcseconds, which is the size of the boxes. }

\label{fig:fov}
\end{figure}

The galaxy cluster M0416 \citep{ebeling, mann} lies at a redshift of
$z=0.396$ and has been extensively studied in recent years due to its
inclusion in CLASH.  While it does not have a particularly large
critical magnification area, it contains a high number of
multiply-lensed images and therefore has one of the most securely
constrained mass models \citep[e.g.,][]{zitrin13, jauzac2014}. M0416
is the second \HFF\ observed cluster in \HST\ cycle 21
(Figure \ref{fig:fov}). The primary \HST\ observations of M0416 span dates
from 2014 Jan. 5 to Sep. 1 (GO/DD~13496, PI: Lotz), while additional
\HST\ observations span dates from 2014 Feb. 1 to Sep. 28 (GO~13386,
PI: Rodney).  ACS and WFC3 images were downloaded from Mikulski
Archive for Space Telescopes
(MAST\footnote{\url{http://archive.stsci.edu/hst}}).  Total exposure
times and limiting magnitudes for each band are listed in Table
\ref{tbl:obs}. We processed the \HST\ data with the {\tt APLUS}
\citep{aplus} pipeline; for details refer to \cite{zheng14}. The
pipeline products include mosaic images, source catalogs and
photometric redshifts.  To complement the \HST\ dataset, deep
\sst/IRAC channel 1 and 2 imaging, which correspond to 3.1--3.9 and
3.9--5.0 \micron, respectively, was acquired between 2013 Dec. and
2014 Mar. using Director's Discretionary Time (Program 90258, PI:
Soifer). Additional archival data (Program 80168; PI: Bouwens) between
2011 Oct.  and 2012 Apr. were also used. The total effective exposure
time per channel is 341 ksec.  Images were processed with the standard
Spitzer pipeline {\tt MOPEX} \citep{mopex} and calibrated as described
in \cite{zheng14}.

Photometry on the \sst/IRAC images is challenging, as many of our
candidates suffer contamination from nearby objects due to the
instrument's large point spread function (PSF, FWHM$\sim$1\farcs6). To
address this issue, we modeled all nearby objects in a 10$\times$10
arcsec$^{2}$ region using the deep F160W band of \HST\ and subtracted
these objects models from the IRAC images using GALFIT \citep{galfit}
to perform relatively accurate aperture photometry. We modeled the
objects with point models for unresolved objects or S\'{e}rsic models for
extended objects convolved with the IRAC PSF. The IRAC PSF was derived
from isolated point sources which were carefully selected in the
same field. The photometry of each candidate was performed with a
2\farcs5 diameter circular aperture after the subtraction of nearby
galaxies. The errors were calculated based on fluctuations in the
residual image. We applied a factor of 2.0 aperture correction to
account for light outside of the aperture in the wings of the
PSF. During this procedure, we occasionally found candidates which a)
lie very close to nearby objects, b) are in crowded regions with
several bright galaxies, c) straddling gaps between different exposure
levels, or d) in regions with complex background; we flag such
candidates as highly contaminated. IRAC photometry for all candidates
are provided in Table \ref{tbl:IRAC}.

\begin{figure}
   \centering
   \includegraphics[width=0.8\textwidth]{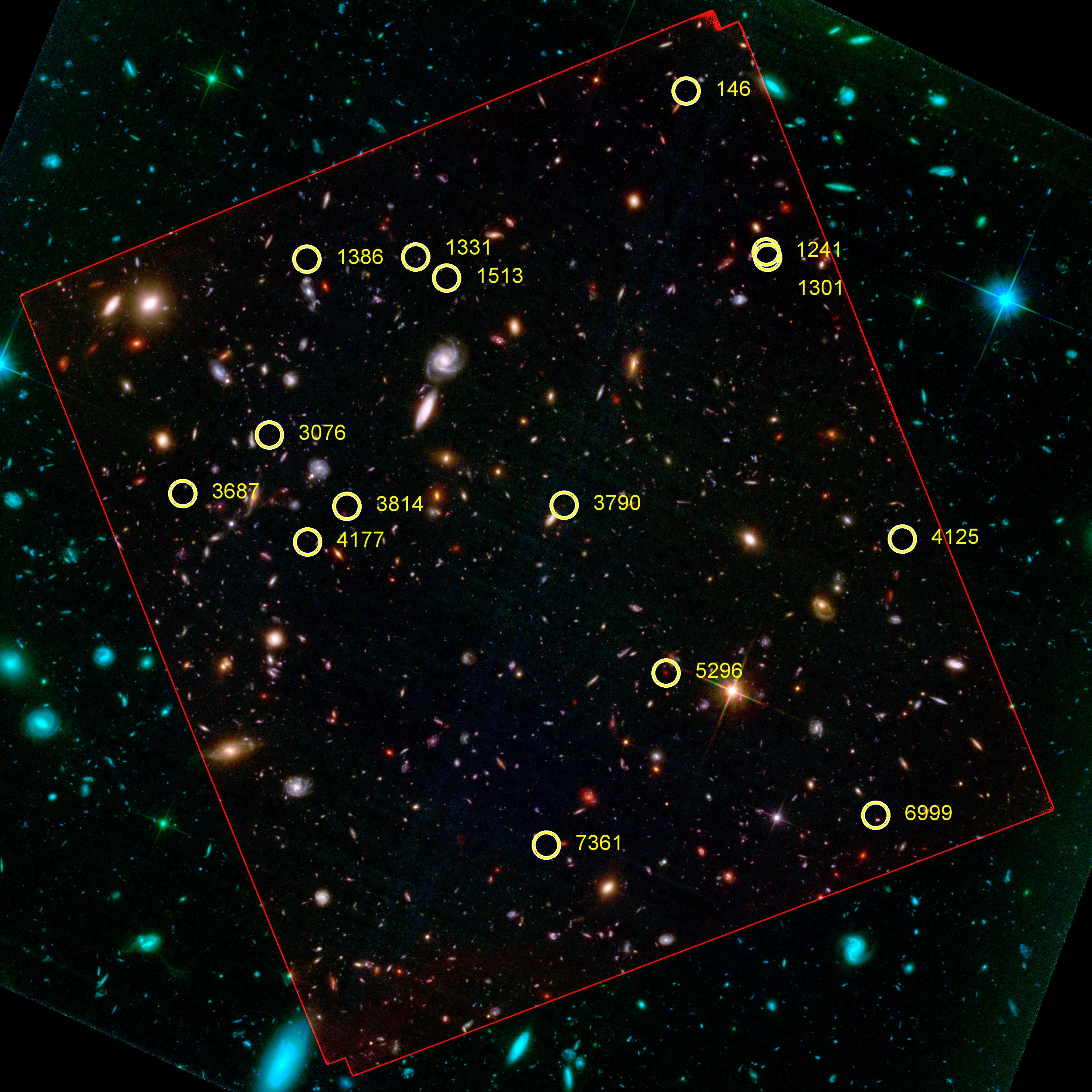}
\caption{A 3\arcmin$\times$3\arcmin \HST\ color image of the M0416
  parallel field, comprised of the ACS F606W (blue), ACS F814W
  (green), and WFC3 F160W (red) bands. Again, our $z>7$ candidates are
  labeled and marked with yellow circles.}
\label{fig:fovp}
\end{figure}

Parallel observations were made during the same period and with the
same filters as the primary \HST\ observations
(Figure \ref{fig:fovp}). Exposure times in these bands are listed in Table
\ref{tbl:obs}. The M0416 parallel field is centered at RA=$04^{\rm h}
16^{\rm m} 33\fs 5$ ~ Dec=$-22\deg 06\arcmin 48\arcsec$.  This field
is also covered by the Spitzer/IRAC observations of M0416.

To increase the wavelength coverage of our survey and therefore 
increase the number of SED constraints for our candidates, we used a 
deep $K_s$ image taken with HAWKI/VLT \citep{pirard} between November 
2013 and February 2014 (ID: 092.A-0472, PI: G. Brammer). The field of 
view of this image covers both the cluster and parallel fields.   
The raw HAWK-I images were processed using a custom pipeline, 
which was originally developed for the NEWFIRM Medium Band Survey 
\citep{whitaker} and later adapted for the ZFOURGE \citep{spitler} 
and HAWK-I Frontier Fields (Brammer, in prep) surveys.  We measured the image 
depth using empty $0\farcs4$ radius apertures distributed over the field, resulting 
in a 5$\sigma$ limiting magnitude of 26.1.

\section{SELECTION}\label{sec:selection}

Candidate selection was performed using typical LBG dropout critieria,
searching for LBG candidates using their distinct color around $0.1216
(1+z)$ \micron. The criteria are as follows: For candidates at
$z\simeq 7$--8, the Lyman break is at $\sim 1$ \micron, between the
F814W and F125W bands: $\I - \Y > 0.8$, $\Y - \J < 0.6$ and $\I - \Y >
0.8 + (\Y - \J)$.  These color cuts are similar to those utilized in
previous work \citep[e.g.][]{oesch10, zheng14}.  For $z\simeq 8$--9,
the break is at $\sim 1.15$ \micron, between the F105W and F140W
bands: $\Y - \JH > 0.8$, $\JH - \H < 0.6$ and $\Y - \JH > 0.8 + (\JH -
\H)$ Finally, for $z\simeq 10$, the break is between the F125W and
F160W bands: $\J - \H > 0.8$.

For each redshift range, a strong requisite is that a candidate should
not be detected above $1\sigma$ in images bluer than the Lyman
break. For $ z \simeq 7 $ objects, a null detection is required in the
synthesized F606W and F435W band, while for candidates at $ z \gtrsim
8 $ a null detection is enforced in a stacked optical image.

In order to avoid contaminated photometry, we further excluded objects
which fall within one arcsecond from the edge of the detector. We also
excluded objects near stellar diffraction peaks. Finally, objects that
have color decrements F160W$ - $IRAC1 $>$ 3 are eliminated, as they
are most likely extremely red objects at lower redshift ($z\simeq 2$).

In addition, we correct to the total flux using $mag_{auto} -
mag_{iso}$ in the F160W band, following \cite{zheng14}. 
 Where blending is a problem, the aperture is visually chosen so that it
will not extend into the other source's lobe.
Furthermore, in those cases where we use the publicly
released \HST\ mosaics, we verify that our aperture colors are not
affected by image artifacts
\citep[e.g.,][]{koekemoer}.\footnote{http://archive.stsci.edu/pub/hlsp/frontier/macs0416/images/hst/v1.0}

\subsection{Photometric Redshift Selection}\label{photoz}

We calculate photometric redshifts (photo-z's) using an updated
version of the Bayesian Photometric Redshifts code \citep[hereafter
BPZ2.0;][]{bpz, coe06}, which includes several changes with
respect to its original version \citep[see][for more
details]{molino}. In particular, we use a new library composed of
six SED templates originally drawn from Projet d'\'Etude des GAlaxies
par Synth\`ese \'Evolutive \citep{pegase} but then re-calibrated using
FIREWORKS photometry and spectroscopic redshifts \citep{fireworks} to
optimize its performance. In addition to these basic six templates,
four GRAphite and SILicate ($GRASIL$) and one $STARBURST$ template
have been added. This new library, already adopted by the CLASH
collaboration \citep{jouvel}, includes five templates for elliptical
galaxies, two for spiral galaxies and four for starburst galaxies,
along with emission lines and dust extinction. The opacity of the
intergalactic medium was applied as described in \citet{madau}.

BPZ2.0 also includes a new empirically derived prior based on the
redshift distributions measured in the GOODS-MUSIC \citep{santini},
COSMOS \citep{scoville} and UDF \citep{coe06} catalogs. However, since
the galaxies considered in this work are (typically) beyond the
redshift range employed to constrain the BPZ2.0 priors, we preferred to
assume an ignorant (i.e., flat) prior on both galaxy type and
redshift, since there is no guarantee that a simple extrapolation to
the high-z Universe may not introduce an unexpected bias in the
analysis.

As already emphasized by several authors
\citep{bpz,coe06,mandelbaum,cunha,wittman,bordoloi,abrahamse,sheldon,carrasco
  kind13,molino,carrasco kind14}, photo-z's should not be treated as
exact estimates, but as probability distribution functions (PDF) in a
bi-dimensional (redshift vs spectral-type; i.e., z-T) space. Although
it is true that for high signal-to-noise detections the PDF can be
well-approximated by a Gaussian distribution, for faint detections the
photometric uncertainties make these distributions highly non-Gaussian
and completely asymmetric. A further difficulty arises when a
sufficiently large wavelength-range coverage is not available (or the
photometry is not deep enough) to simultaneously identify at least two
distinct prominent spectral features (as is usually the case for the
detection of very high-z galaxies), enabling more than one solution to
fit the input photometric data equally well. This problem, known as
the color-redshift degeneracy, makes the PDF bimodal since the
4000\AA-break from an early-type galaxy at low-redshift cannot be
distinguished from the Lyman-break of a late-type galaxy at
high-redshift. This high-to-low redshift misclassification problem is
of major importance in the identification of high-z candidates.

\begin{figure}[h]
   \includegraphics[width=0.8\textwidth]{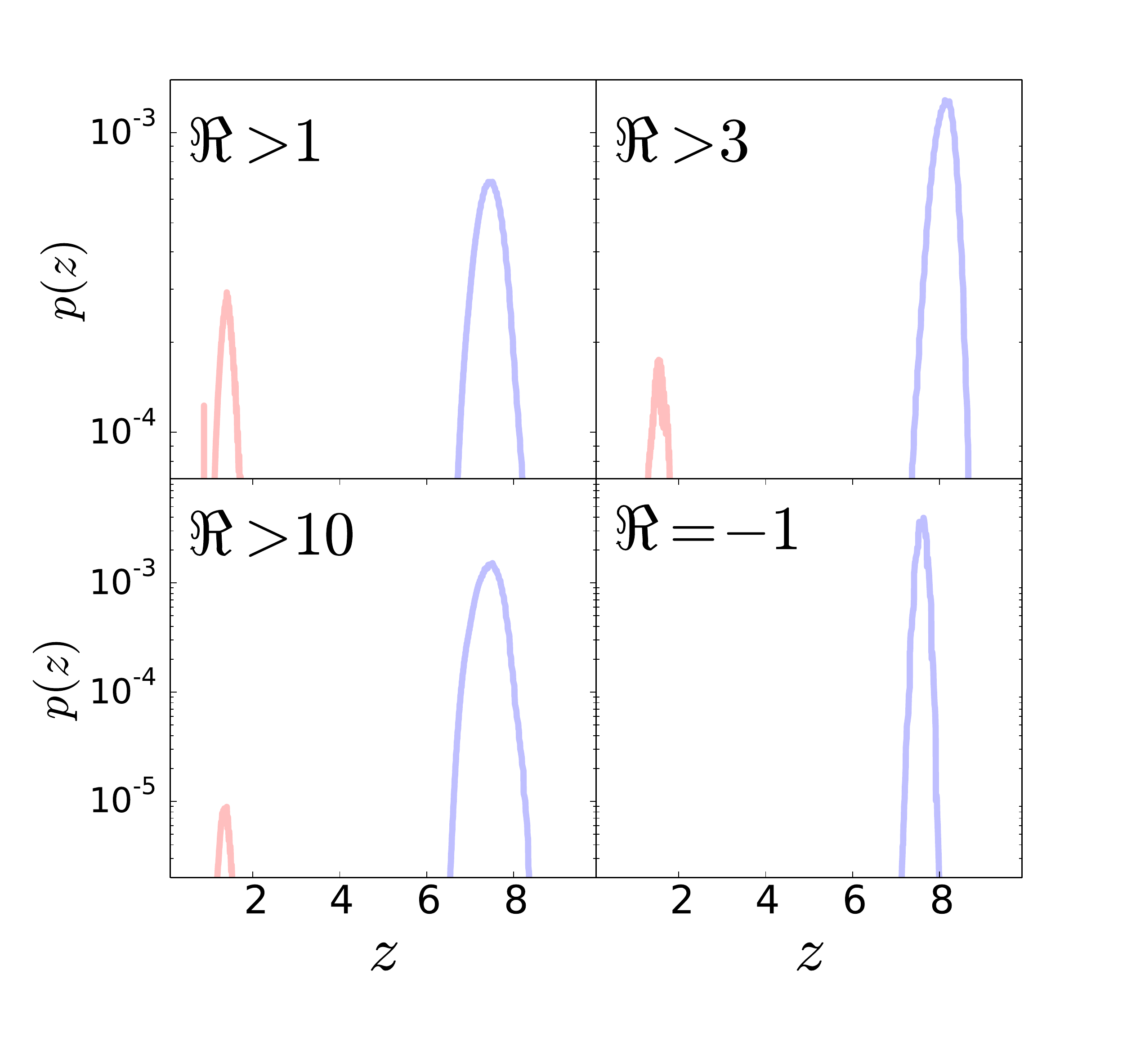}
\caption{The color-z degeneracy problem and its separation among
  spectral-types. For the identification of high-z galaxies, if a
  sufficiently large wavelength coverage is not available or the
  photometry is not deep enough for the simultaneous identification of
  the most prominent spectral features, several solutions may equally
  fit the input photometric data. This problem, known as the color-z
  degeneracy, makes the PDF bimodal since the 4000\AA-break from an
  early-type galaxy at low-z cannot be distinguished from the
  Lyman-break from a late-type galaxy at high-z. However, since this
  bimodality represents two independent scenarios for a single galaxy
  (left: early-type/low-z or right: late-type/high-z), the global PDF
  can be separated in two independent redshift distributions, allowing
  the probability comparison of both scenarios. We show four
  examples for $\Re = 1,3,10,-1 (\infty)$.}
\label{fig:pdz}
\end{figure}

However, since this bimodality represents two independent scenarios
for a single galaxy (early-type/low-z or late-type/high-z, see figure
\ref{fig:pdz}), the global PDF can be easily separated into two
independent redshift distributions, as expressed in equation
\ref{phzeq1}:

\begin{eqnarray}
\label{phzeq1}
p(z) = \int p(z,T) \, dT=   \,\,\,\,\,\,\,\, \\
= \int p(z,T_{E}) \, dT + \int p(z,T_{L}) \,dT  \nonumber
\end{eqnarray}

This simple separation between ``red'' templates ($T_{E}$) and
``blue'' templates ($T_{L}$) renders it possible to make simple
comparisons between both scenarios. Following the same philosophy used
by \citet{lopez-sanjuan}, we express the normalized global PDF for
every galaxy in terms of the probability of having an early ($p_{E}$)
or a late ($p_{L}$) spectral-type solution, as indicated in equation
\ref{phzeq2}:

\begin{eqnarray}
\label{phzeq2}
\mathrm{P(z)} = \int p(z) \,dz = \mathrm{p_{E}(z)}  + \mathrm{p_{L}(z)} = 1
\end{eqnarray}

Likewise, we define (eq. \ref{phzratio}) the peak-probability ratio
$\Re$ as the ratio between peak-probabilities; i.e., between the
values that each distribution takes as its peak (or maximum), such
that:

\begin{eqnarray}
\Re \equiv \frac{p_{L}(z=z_{peak,L})}{p_{E}(z=z_{peak,E})}
\label{phzratio}
\end{eqnarray}

\begin{figure}[h]
   \centering
   \includegraphics[width=0.8\textwidth]{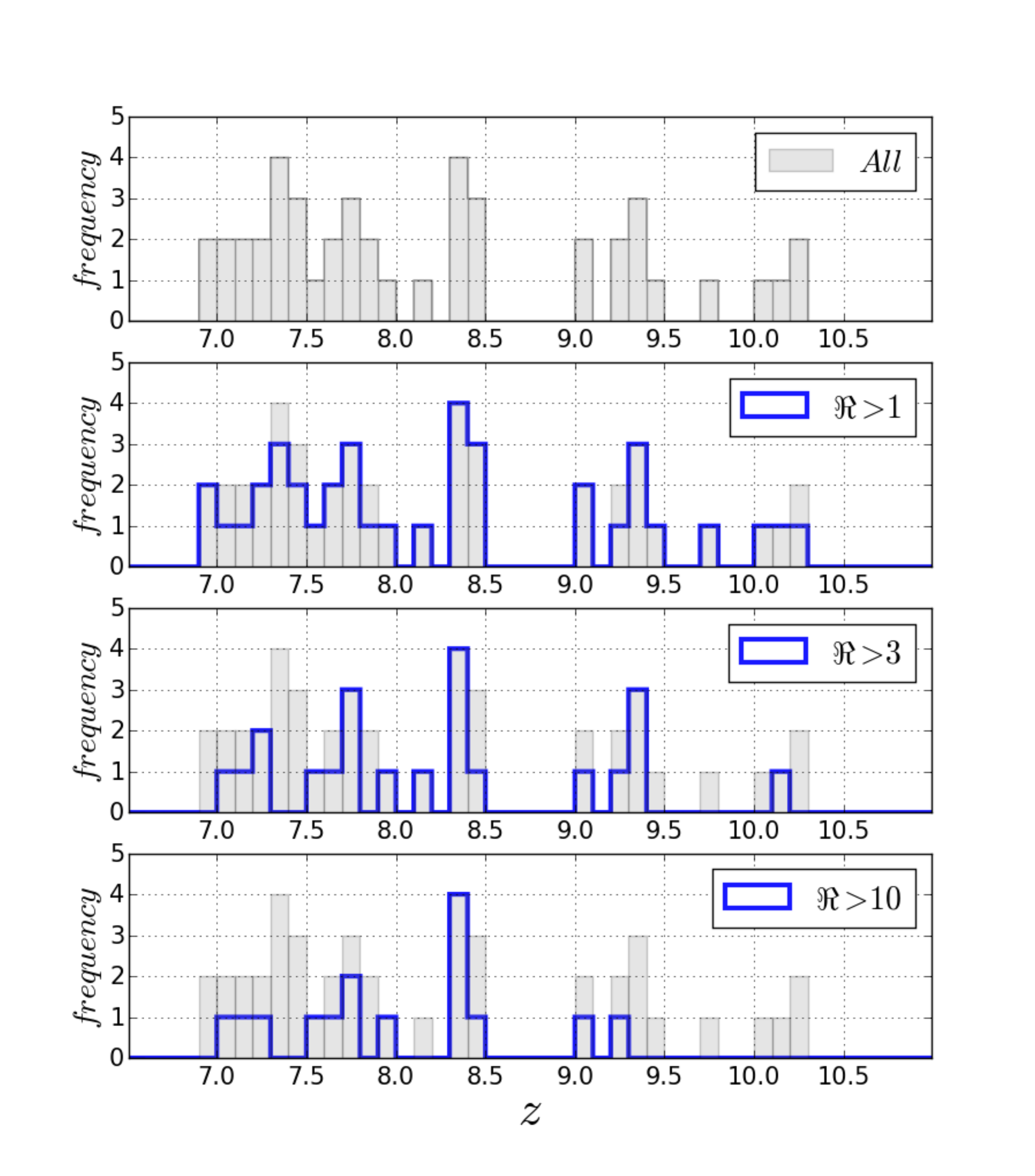}
\caption{Distribution of peak photometric redshifts for entire sample
  of the $z>7$ candidates from the M0416 cluster and parallel
  fields. Once the global PDF is separated among spectral-type
  solutions, we define the peak-probability ratio ($\Re$) as the ratio
  among peak-probabilities; i.e., between the values that each
  distribution takes in its peak (or maximum). This number roughly
  quantifies how many times one solution is more likely than the
  other, and thus allows to flag potential z$\gtrsim$7 candidates.
  The first panel shows the distribution for the entire sample of 45
  photometrically selected galaxies. From these, a total of 15
  ``potential'' candidates have a minimum threshold of
  $\Re>10$, 22 candidates have $\Re>3$ and 38 candidates have $\Re>1$.
  $\Re>3$ photometric redshifts are used in our analysis.}
\label{fig:zhist}
\end{figure}

This number approximates how many times one solution is more likely
than the other, and so makes possible to quantitatively flag potential
high-z candidates. As shown in Table ~\ref{tbl:photoz}, where both the
photometric redshift estimations and the peak-probability ratios are
presented for every photometrically selected candidate, we find a
total of 15 ``potential'' candidates lie at z$\gtrsim$7 if we impose a
minimum threshold of $\Re>10$. If this condition is relaxed to $\Re>3$
or $\Re>1$, we obtain 22 and 38 candidates, respectively. The
resulting redshift distribution (when applying the different threshold
criteria) is shown in Fig. \ref{fig:zhist}. To be on the conservative
side, we consider as high probability only those high-redshift
candidates with $\Re>3$. Moreover it is interesting to note that the
use of $\Re>3$ instead of $\Re>1$ suggests a contamination rate of
$\approx$40\%, in the order of what is found in previous surveys
 (e.g. \citealt{bradley12}, \citealt{Schmidt}).  We know the peak ratio $\Re$ 
is probably not the best estimate, it is just the simplest and \textit{good enough}
for our purpose.

\begin{figure}
   \centering
   \includegraphics[width=1.0\textwidth, angle =0]{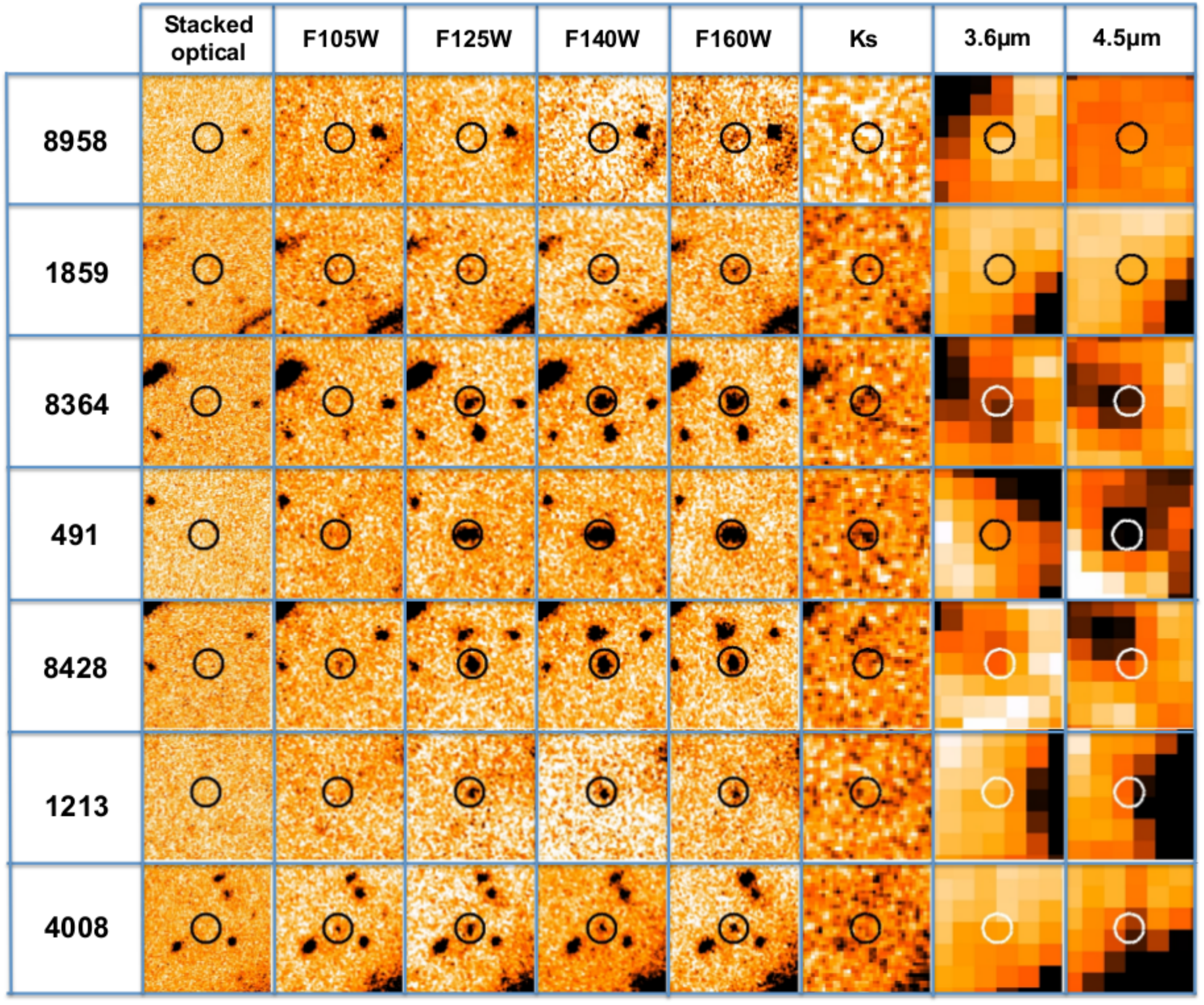}
\caption{Postage stamp images for all $z>7$ candidates in the
  M0416 cluster field. The stamp size is 4\arcsec$\times$4\arcsec and
  the radius of the circle is 0\farcs4.}
\label{fig:stamps-c}
\end{figure}

\begin{figure}
   \centering
   \includegraphics[width=1.0\textwidth, angle =0]{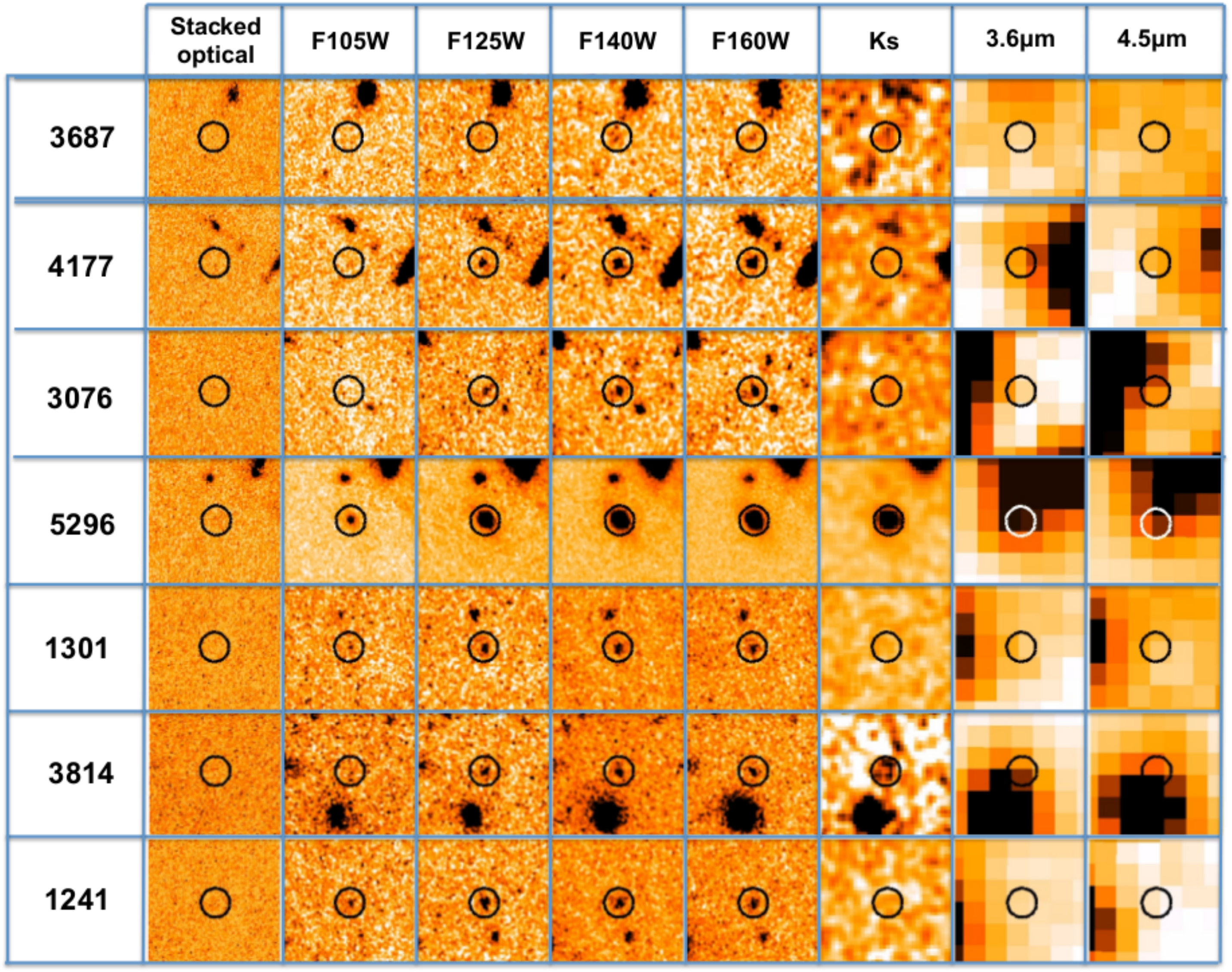}
\caption{Postage stamp images for all $z>7$ candidates in the M0416
  parallel field, part 1. Details are the same as
  Figure~\ref{fig:stamps-c}.}
\label{fig:stamps-p1}
\end{figure}

\begin{figure}
   \centering
   \includegraphics[width=1.0\textwidth, angle =0]{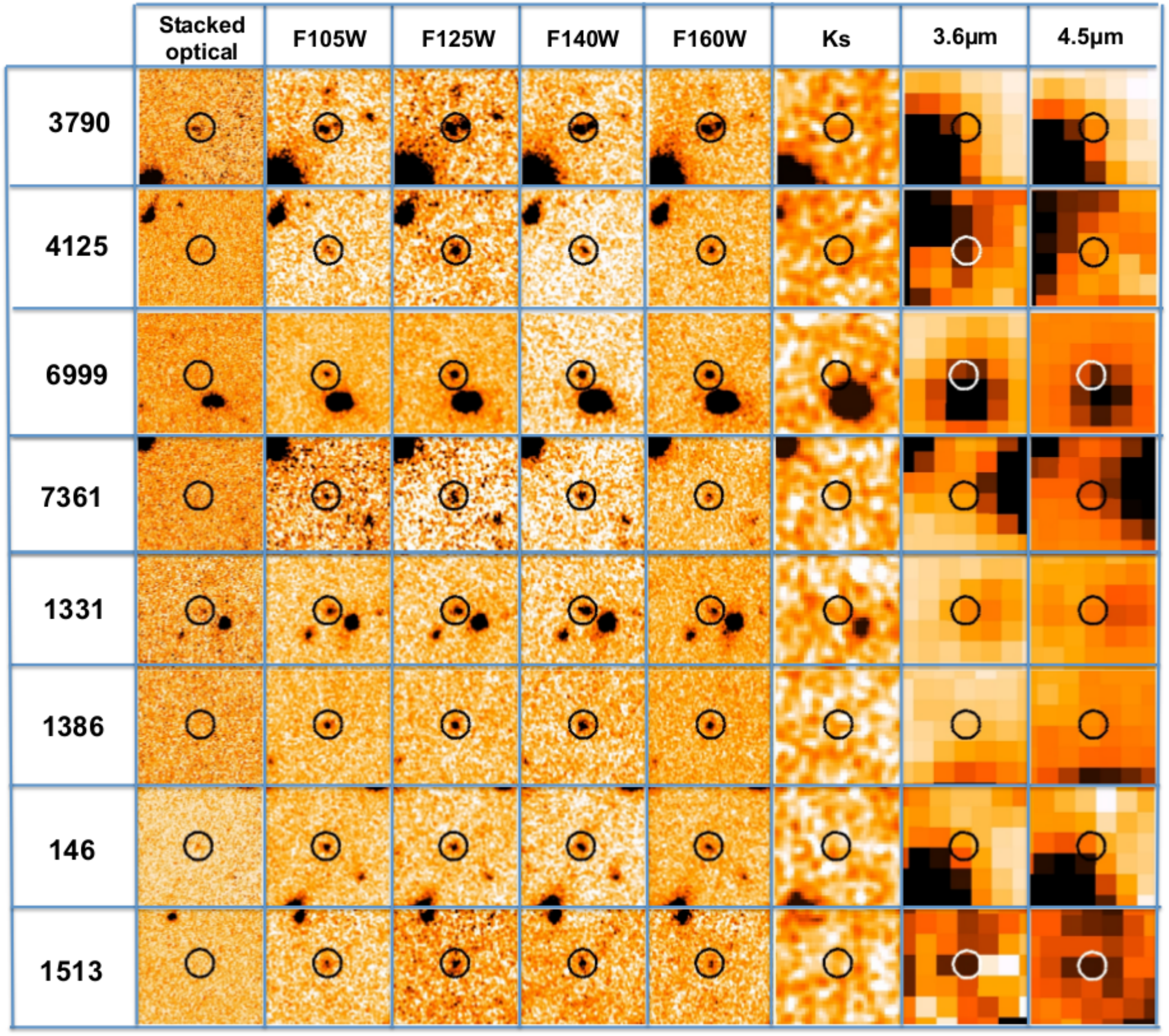}
\caption{Postage stamp images for all $z>7$ candidates in the M0416
  parallel field, part 2. Details are the same as
  Figure~\ref{fig:stamps-c}.}
\label{fig:stamps-p2}
\end{figure}

In Tables~\ref{tbl:z7p}-\ref{tbl:z7} we present identifications,
positions and photometry for all our selected candidates with
$\Re>3$. In Figures~\ref{fig:stamps-c}, \ref{fig:stamps-p1}, and
\ref{fig:stamps-p2}, we show cutout images of all the $\Re>3$
candidates.  In the discussion that follows we only
consider these candidates and use these BPZ based redshifts for all further 
calculations, e.g. luminosity function, physical properties, etc.

\section{LENS MODELS}\label{sec:lens}

We derive magnifications for the different candidates from a
selection of lens models publicly available through the MAST
archive\footnote{http://archive.stsci.edu/prepds/frontier/lensmodels/}.
In order to maintain consistency among these models, we select only
those that incorporate the cluster members into the lens model.

In particular, we use the two models delivered by the {\it
Sharon}-team \citep{Johnson14}, the two models delivered by {\it
Zitrin \& Merten}-team \citep{zitrin13} and one model delivered
by the {\it CATS}-team \citep{richard14}.  In addition to these five
models, we include also the free-form model from \cite{diego15} which,
as the models above, includes the contribution from cluster members to
the lens model. 

Based on these six models we compute the mean magnifications
at the corresponding redshifts for the candidates listed in Table~\ref{tbl:z7} as well as the 
dispersion in the magnification between the different models. 
The candidates that are further away from the
critical curves show smaller dispersion between the predicted magnification values. 
The critical curves for one of these models \citep{diego15} together with 
the position of the candidates around the central part of the cluster are shown in Figure \ref{fig:fov}.

For candidates near the critical curves, we use the model to predict
potential counter-images.
Variations in the predictions of counter images are expected between different
models specially near the critical curves. For candidate 8958 we find that the 
models predict 2 additional counter-images. The precise location varies depending 
on the particular model but with a fair agreement between different models. 
Based on the \cite{diego15} model, counter images should be found at a) 
RA=64.023524 $deg$, DEC=-24.07943 $deg$ ($\mu=3.9$) and at b) RA=64.036003 $deg$, 
DEC=-24.08669 $deg$ ($\mu=1.7$). At both positions, we locate several faint 
red objects within 2 arcseconds of the predicted position. In particular, for 
the predicted position a) we find 3 faint candidates within 1.5 arcseconds 
and for the predicted position b) we find one candidate $\approx$ 1.8 
arcseconds away. However, despite the presence of possible candidates around 
the predicted positions, we note that the relative weaker magnification 
at the predicted positions compared with the large magnification for the already 
faint candidate 8958 ($\mu \approx 20$) makes it difficult to reach a reliable 
identification of candidates at the predicted positions.

\section{DISCUSSION}\label{sec:discussion}

\subsection{Contamination}

One possible source of contamination to the high-$z$ sample is from low-$z$ interlopers (e.g. \citealt{Hayes12}) that could fulfill all the selection criteria defined for very high-redshift objects. We estimated the contamination by this type of source by simulating a population of galaxies based on the distribution of galaxies in our dataset. We first build this "contaminant sample" by selecting all objects that are brighter than the brightest candidate selected in the field (F160W$>$25.5) and that are detected at more than 2$\sigma$ in optical bands (for the parallel field we used the same limit even though we selected one object at F160W=24.18, see discussion below). We then matched the luminosity distribution of the bright objects to the luminosity of our candidates. We measured the noise around each object to update the photometry of the contaminants sample. Finally we applied the same selection criteria as those we used for the high-$z$ selection: all the objects that enter the selection window are contaminants. As a first step, we only applied the classical LBG selection criteria (non-detection and color-selection) and found a contamination rate of $\sim$28\% for the cluster sample and $\sim$42\% for the parallel sample. We then applied to the selected contaminants the peak probability ratio method and found that all the objects in the contaminants sample display a ratio R $<$ 2, such that they are not selected as good candidates. 

Moreover, the high-$z$ sample can also be contaminated by M, L and T dwarfs exhibiting colors similar to what is required for high-$z$ galaxies. We used observed spectra of such stars published in \citet{burgasser04} and \citet{burgasser10} and showed that this kind of contaminant has the following colors : F105W-F140W$<$1.0 and F140W-F160W$<$0.4. Therefore, among our samples, 9 objects are consistent with these colors, and require further investigation. We measured the size of these nine objects following the method described in \citet{oesch10} correcting for psf broadening and amplification, and demonstrated that all 9 objects are resolved on HST images.

\subsection{The brightest $z\sim$8 candidate}

Among our two samples, one $z\sim$8 candidate appears extremely bright (F140W $\sim$ 25.5) without being strongly amplified because it is located in the parallel field (\#5296). The expected number of such bright objects in the full Frontier Fields dataset ($\sim$32 arcmin$^2$ according to \citealt{coe15}) assuming the evolution of the UV luminosity function found by \citet{bouwens4} is ($1.33^{+5.00}_{-0.96}$)$\times$10$^{-3}$. We also note that only one object with F140W$<$ 24.5 is expected in the final release of the UltraVISTA survey \citep{UltraVISTA}, implying that a high-redshift hypothesis is unlikely. In order to asses whether this extreme object is an interloper or a bona fide high-redshift galaxy, we perform several tests.

We first study the non-detection in the optical bands by applying the $\chi^2_{opt}$ method defined in \citet{bouwens11}. We compare the distribution of $\chi^2_{opt}$ measured in empty apertures all over the field with the distribution of $\chi^2_{opt}$ for mock objects detected at $\sim$2$\sigma$ in all ACS bands, and show that all sources with $\chi^2_{opt}>$0.4 are most likely contaminants. The $\chi^2_{opt}$ of our source is in perfect agreement with the $\chi^2_{opt}$ measured in empty apertures [$\chi^2_{opt}$(\#5296)=-0.04].  However, we notice a possible faint detection of that object on the F814W image ($>$30.1) that could make unlikely the $z\sim$8 hypothesis, but that confirms a huge break between optical and NIR data of at least 4.5 mag. 

The star hypothesis was studied by measuring the size of this object and by comparing its colors with expected colors of M, L and T dwarfs, that could be similar to those of very high-$z$ objects. We measured its size using the SExtractor FLUX\_RADIUS parameter which encloses 50\% of the light, as described in \citet{oesch10}, and corrected this value for PSF broadening. This shows that this object is clearly resolved, and assuming that this object is at $z\sim$ 8.36, its physical size would be 0.55 kpc, which is consistent with the observed evolution of size as a function of UV luminosity (e.g. \citealt{laporte2}, \citealt{kawamata}). 
As in $\S$5.1, we compared the colors of this candidate (F105W-F140W=1.7 and F140W- F160W=0.31) to those expected from M, L and T dwarfs (F105W-F140W$<$1.0 and F140W-F160W$<$0.4), finding that they are incompatible.

The last hypothesis we study for this target is the possibility that it is a moving object. We took benefit from the 4 epochs during which the parallel field has been observed with WFC3/HST, which span over 1 month. We ran SExtractor on each epoch image and compared the centroid position for that object; it shows no evidence of movement.

\subsection{Physical Properties}\label{sec:pp}

Next, we characterize the physical properties of our
high-redshift candidates by means of the Bayesian spectral energy
distribution (SED) modeling code {\tt iSEDfit} \citep{moustakas}.
Using the same setup and parameter set as used in
\cite{zheng14}, we generated 80,000 models SEDs with \emph{delayed}
star-formation histories using a Monte Carlo technique.  iSEDfit also
includes nebular emission lines. The SEDs were
computed employing the Flexible Stellar Population Synthesis models
\citep[FSPS, v~2.4;][]{conroy09, conroy10} based on the {\sc miles}
stellar library \citep{sanchez-blazquez06}, and assume a
\citet{chabrier} initial mass function from $0.1-100~M_{\odot}$.

We adopt uniform priors on stellar metallicity
$Z/Z_{\odot}\in[0.04,1.0]$, galaxy age $t\in[0.01,age(z_{BPZ})]$~Gyr,
star formation timescale $\tau\in[0.01,age(z_{BPZ})]$~Gyr, where
$age(z_{BPZ})$ is the age of the Universe at each galaxy's
photometric-redshift, and we assume no dust attenuation \citep[{\it
    e.g.},][]{bouwens3}.  In Table \ref{tbl:pp} we report the median values of
the posterior probability distributions for some physical properties
and their 1-$\sigma$ confidence level uncertainties.  
Our $z>7$ candidates have median stellar masses of
$log(M_*)\sim 8.40^{+0.55}_{-0.31}$~\Msun, SFRs of approximately
$1.6^{+0.5}_{-0.4} $~\Msun~yr$^{-1}$, and SFR-weighted ages of
$\lesssim 310^{+70}_{-140}$~Myr.

\begin{figure}
   \centering
   \includegraphics[width=1.0\textwidth]{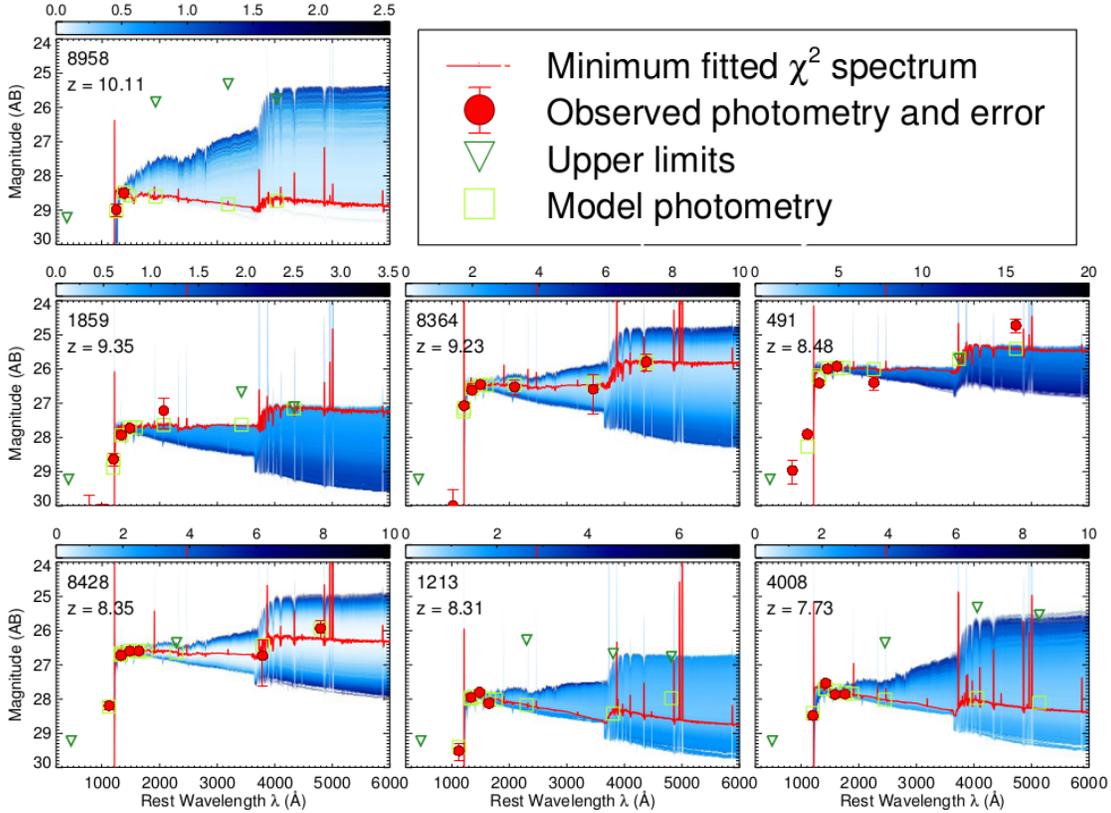}
\caption{Rest-frame SEDs of the $z \geq 7$ candidates in the M0416
  cluster field. Filled red circles denote observed ACS, WFC3,
  HAWKI/VLT and IRAC photometric detections, while open green
  triangles indicate $3\sigma$ upper limits.  The red spectrum shows
  the best-fitting (maximum likelihood) SED template based on our
  Bayesian SED modeling using {\tt iSEDfit}.  The large green squares
  show the expected photometric magnitudes of the best-fitting model
  convolved with the ACS, WFC3, and IRAC filter response curves. 
The blue shading shows some models, which were used in the fitting process, 
scaled by their $\chi^2$. The colorbar shows the $\chi^2$ scale.
}
\label{fig:SED_m0416}
\end{figure}

\begin{figure}
   \centering
          \includegraphics[width=1.2\textwidth]{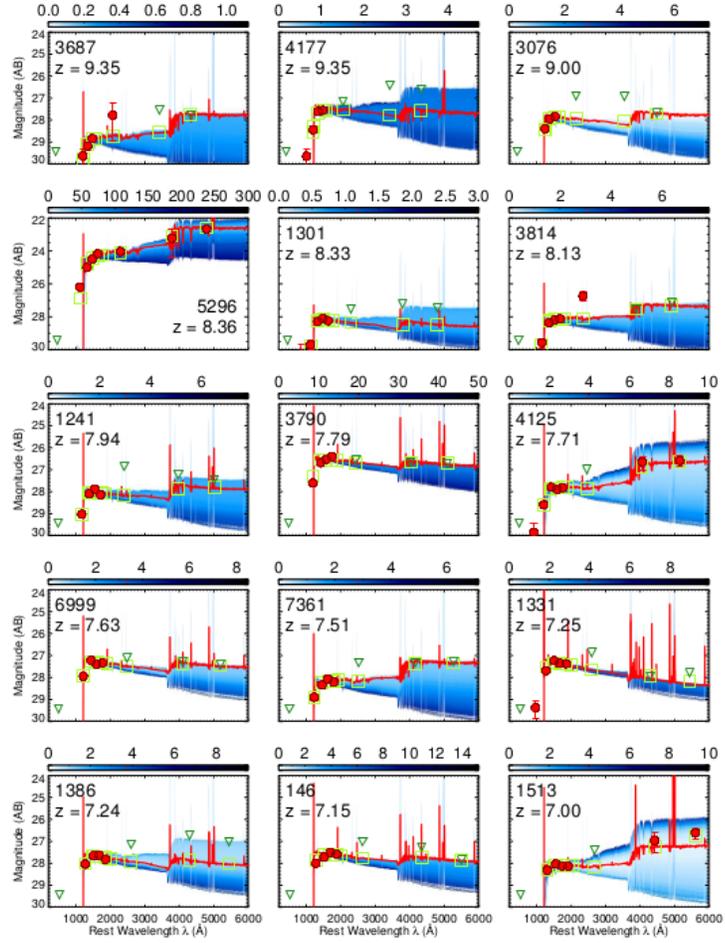}
\caption{Rest-frame SEDs of $z \geq 7$ candidates in the parallel field. Details are the same as
Figure~\ref{fig:SED_m0416}. }
\label{fig:SED_p}
\end{figure}

Figures~\ref{fig:SED_m0416} and ~\ref{fig:SED_p}
present the SEDs of the high-z candidates in the M0416 cluster and
parallel fields, sorted by decreasing redshift.  

In order to evaluate the validity of the ``no-dust'' condition adopted
above, we also create 40,000 models employing the time-dependent
attenuation curve of \cite{charlot00} and rest-frame $V$-band
attenuations of $A_{V}\in[0-3]$~mag.  We note that our SEDs are
generally best-fitted with the models having little or no dust
attenuation, obtaining a median $A_{V}$ of $0.28^{+0.16}_{-0.23}$~mag,
and  a similar median stellar mass
[$log(M_*)\sim 8.53^{+0.36}_{-0.29}$ and age $\sim 210^{+60}_{-130}$~Myr]
as those reported previously without considering dust attenuation.
Yet, dust has a big impact on the SFRs, 
its median value increases by a factor of two 
from $1.6^{+0.5}_{-0.4} $~\Msun~yr$^{-1}$ to 
$\sim 3.5^{+0.5}_{-0.2}$~\Msun~yr$^{-1}$ . The reported median
values are based on the median values of each posterior probability
distribution.

 Finally, since the physical-property estimation depends 
on the photometric redshift assumption, we also use {\tt iSEDfit }
to calculate photometric redshifts. The comparison of these 
results allow us to evaluate how reliable is our estimation 
of the physical properties based on BPZ redshifts.
We generate 40,000 SEDs models adopting the same SSP models, stellar
libraries, parameterization of the star formation history, and IMF as
used in the previous calculation of the physical properties including
dust (\cite{charlot00} curve and $A_{V}\in[0-3]$~mag).  Nevertheless,
in order to additionally encompass possible low-redshift interlopers,
we model these SEDs within a broader range of uniform priors on age
$t\in[0.01,13.5]$~Gyr, star-formation timescale
$\tau\in[0.01,5.0]$~Gyr, stellar metallicity
$Z/Z_{\odot}\in[0.04,1.6]$ than previously. Despite the fact that {\tt
  iSEDfit} was optimized to estimate the physical properties and not
photometric redshifts, we note that the photo-z's obtained with {\tt
  iSEDfit} and BPZ2.0 are reassuringly similar.  {\tt iSEDfit} tends to
recover slightly higher photo-z's compared to BPZ2.0, although the values
are consistent within the error bars, yielding a mean difference of
$\Delta z=0.1\pm0.1$ ({\tt iSEDfit} minus BPZ2.0). 
Since there are no substantial differences between 
the photo-z's based on BPZ and iSEDfit, 
except for the candidate 5296, we can rely on the 
physical properties estimation based on the BPZ 
redshift. It excludes the outlier high-redshift candidate 5296, 
where both photometric redshifts clearly differ, 
$z(BPZ2.0)=8.360^{+0.011}_{-0.004}$ and z({\tt iSEDfit})=$6.9\pm0.4$. 
Hence in Table 6, we report the physical properties according 
both photometric redshifts for this particular candidate.

These physical properties agree with the results of \cite{schaerer10}.
The relatively small size of our sample and their faintness does not allow us to
draw a firm conclusion regarding a star-forming main-sequence is in place at z$>$6. 
Nevertheless, we do confirm that our candidates do conform to the reddening-stellar-mass plane proposed by \cite{schaerer10}: $A_V = log_{10}(M/10^{8} Msun)^n$ with $n=0.4-0.7$.

\subsection{Number Densities at $z\sim$7, 8, 9 and 10}\label{sec:LF}

 The deep images obtained from Frontier Fields observations and the amplification of the light coming from background sources by the cluster allow probe a large range in intrinsic luminosities, thereby allowing us to constrain the UV luminosity function
to unprecedentedly faint levels.  We use the two samples selected in the cluster and parallel fields to compute the number densities in the redshift range covered by our survey. We apply a Monte Carlo method based on the redshift probability distribution \citep{laporte2}. 
We summarize the 8 steps we use to estimate the shape of the UV LF : 
\begin{itemize} 
\item For each object in our samples, we compute the cumulative redshift probability distribution, $P_{cum}(z)$.
\item We choose a random probability, $a$, that assigns a redshift, $z_a$ for each object from the cumulative redshift probability distribution such that $P_{cum}(z_a)$=$a$
\item We use this redshift and the SED of each object to estimate their UV Luminosity corrected for amplification.
\item Steps 1 and 2 are repeated $N$ times in order to get a sample with $N$ times the size of the original sample, but with the same redshift distribution. We adopt a value $N$ large enough so as to not affect the final result (in our case $N=10000$)
\item We distribute all the objects with a redshift included between $z-0.5$ and $z+0.5$, where $z=7, 8, 9$ or 10, in magnitude bins and correct each object for incompleteness. 
\item Each number is corrected by the contamination rate.
\item We compute the effective surface explored by the MACS0416 dataset by matching the amplification map with the detection picture, and then divide the number of objects per magnitude bin by the comoving volume explored between $z-0.5$ and $z+0.5$, where $z=7, 8, 9$ or 10. 
\item Uncertainties on each density are computed using the method described in \citet{cv}
\end{itemize}

The number densities are then corrected for incompleteness due to the extraction methods and the selection criteria we apply to catalogs. The selection function we use to build our high-z samples involves the "standard" Lyman Break technique \citep{steidel}  and the use of the P(z) ration between early type and late type galaxies. Therefore we divided the completeness of our sample into two parts, corresponding to each selection criteria we used, as follows :
\begin{equation}
\label{eq.compl}
C_{LBt}(m,z) \times C_{pic}(m,z)
\end{equation}
The first part of the previous equation, $C_{LBt}(m,z)$, is the incompleteness due to the Lyman Break technique and more especially to the extraction method and color criteria we used. To estimate the shape of this function, we simulate 680,000 galaxies from the latest Starburst99 library (\citealt{leitherer}, \citealt{leitherer10}, \citealt{leitherer12}, \citealt{vazquez}) and other theoretical templates (\citealt{Bruzual and Charlot}, \citealt{coleman}, \citealt{kinney}, \citealt{polletta} and \citealt{silva}). The space parameters we used had magnitudes ranging from 22.00 to 31.00 and redshift from 6.5 to 10.5. We then defined an object mask from the detection picture to identify positions where no object is detected and then added the simulated objects to the real data, without masked objects. We assumed a log-normal distribution of the size of very high-redshift objects with a mean value of 0.15'' and a sigma of 0.07'' as expected from previous studies (e.g \citealt{oesch10}). We then used the same extraction software and selection criteria as defined in section \ref{sec:selection} and compared the number of extracted objects with the number of simulated objects. 

The second part of Eq.\ref{eq.compl}, $C_{pic}(m,z)$, is computed from a mock catalog containing $\approx$50000 objects with redshift ranging from 0 to 12 and a luminosity distribution matched to the $z\sim$6 luminosity function published in \citet{bouwens4}. We run BPZ2.0 and computed for each object the ratio between the probability for this object to be an Early Type (low-$z$) galaxy and the probability to be a Late Type galaxy (high-$z$). The completeness is computed by applying the P(z) selection criteria to this mock catalog and by comparing the selected sample with the number of high-$z$ simulated objects.

The number densities we found are reported on Table \ref{tbl:densities}. It is
interesting to note that one object at $z\sim$10 is strongly amplified
($\mu\sim$20) and thus allows us to give for the first time a robust
constraint in the faint-end ($L\sim0.04L^*$) of the UV LF at $z\sim$10. This result is
consistent with the shape of the LF observed at the brightest
luminosities at such high-redshift.

Recently several papers have demonstrated that the role of foreground
galaxies in the amplification of the light coming from very high-redshift
objects is negligible ($<$1\%) for objects selected in HST surveys
(e.g. \citealt{fialkov}). Therefore we only corrected photometry
for lensing by galaxies cluster in both fields. 

One also has to bear in mind that lensing itself introduces incompleteness to the reconstructed luminosity function. While the so called magnification bias (Broadhurst et al. 1995), which preferentially biases towards the detection of more magnified objects in lensed fields, is effectively folded in our derivation of the LF, there are other lensing-related biases that need to be addressed. For example, Oesch et al. (2014) have shown that both shear (or the apparent change in shape of lensed images) and source blending, have a noticeable effect on the reconstructed LF, biasing the derived SFRD by order $0.4(\pm 0.4)$ dex. Robertson et al. (2014), have shown that cosmic variance -- because of the smaller source plane area being effectively probed with lensing -- is higher and becomes quite significant in lensed fields. In that sense, LFs compiled from a single lensed field may be substantially biased compared to the blank field LFs. Fialkov \& Loeb (2015) show that particularly for Lyman-break high-redshift galaxies, incompleteness from various factors including also the surface-brightness slope of the lensed galaxies, is important to account for (for other discussions of such effects see also Wong et al. 2012, Atek et al. 2014, Ishigaki et al. 2014).

\begin{figure}
   \centering
           \includegraphics[width=4.5cm,angle=-90]{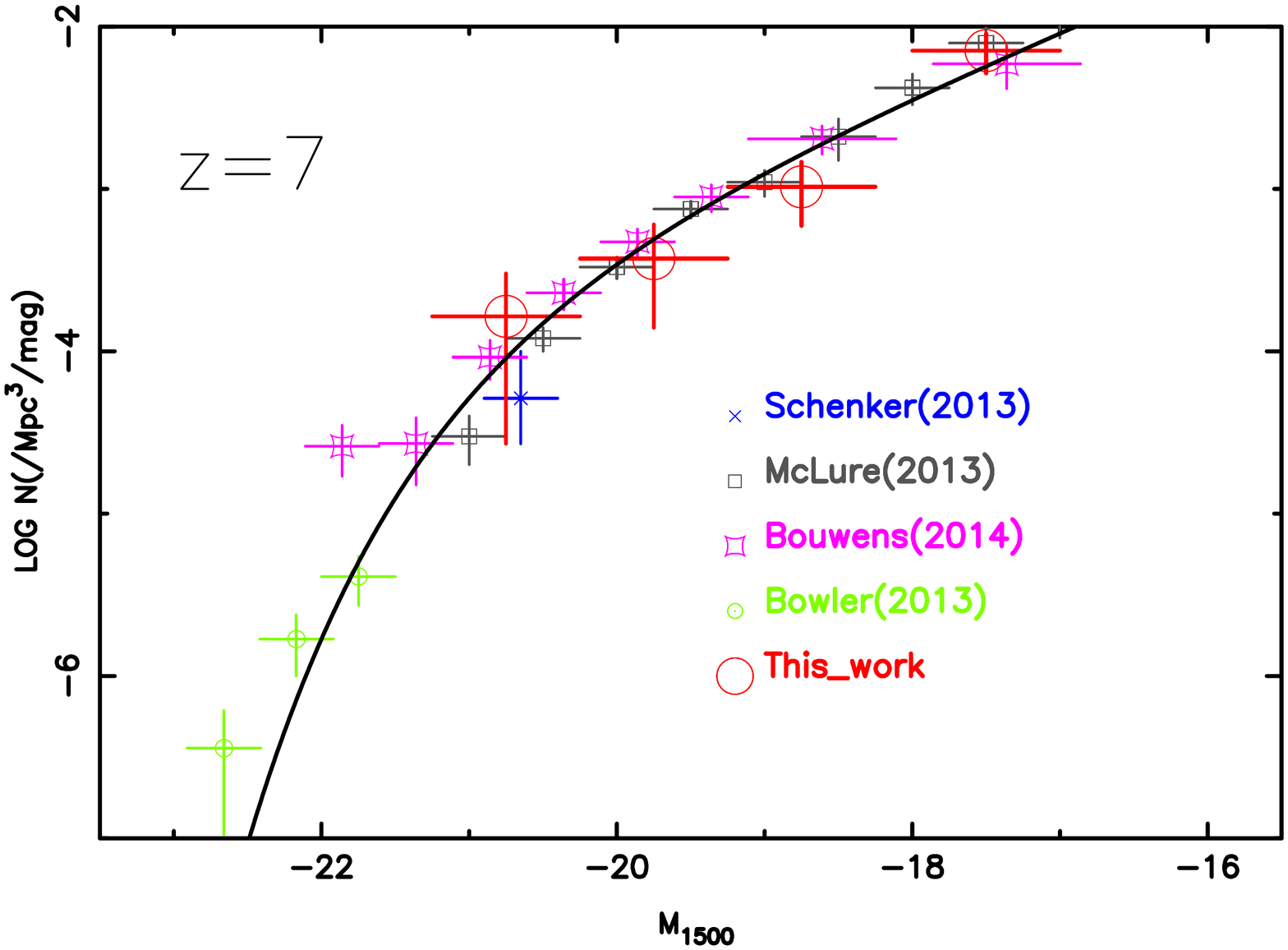}
           \includegraphics[width=4.5cm,angle=-90]{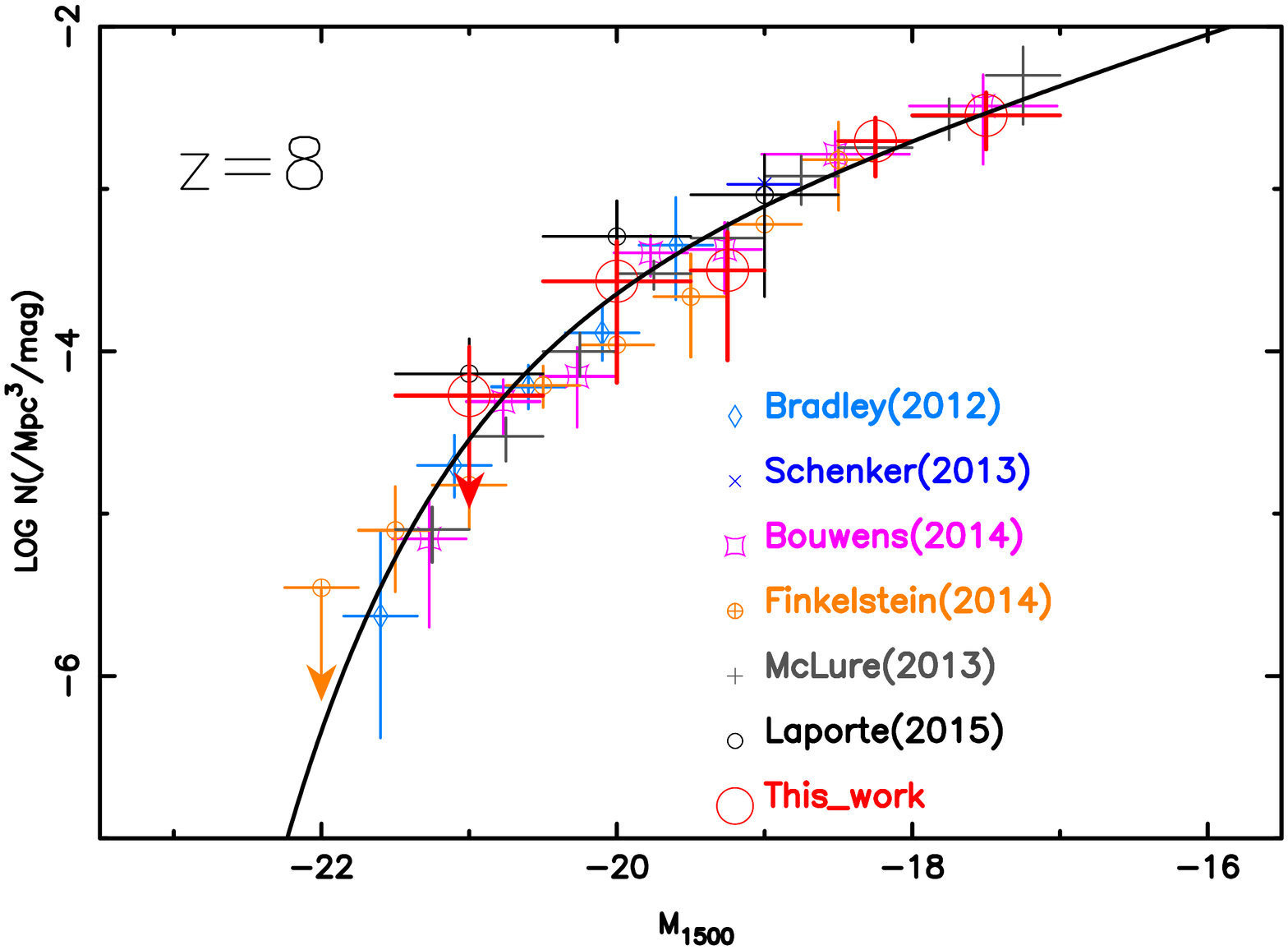}\\
            \includegraphics[width=4.5cm,angle=-90]{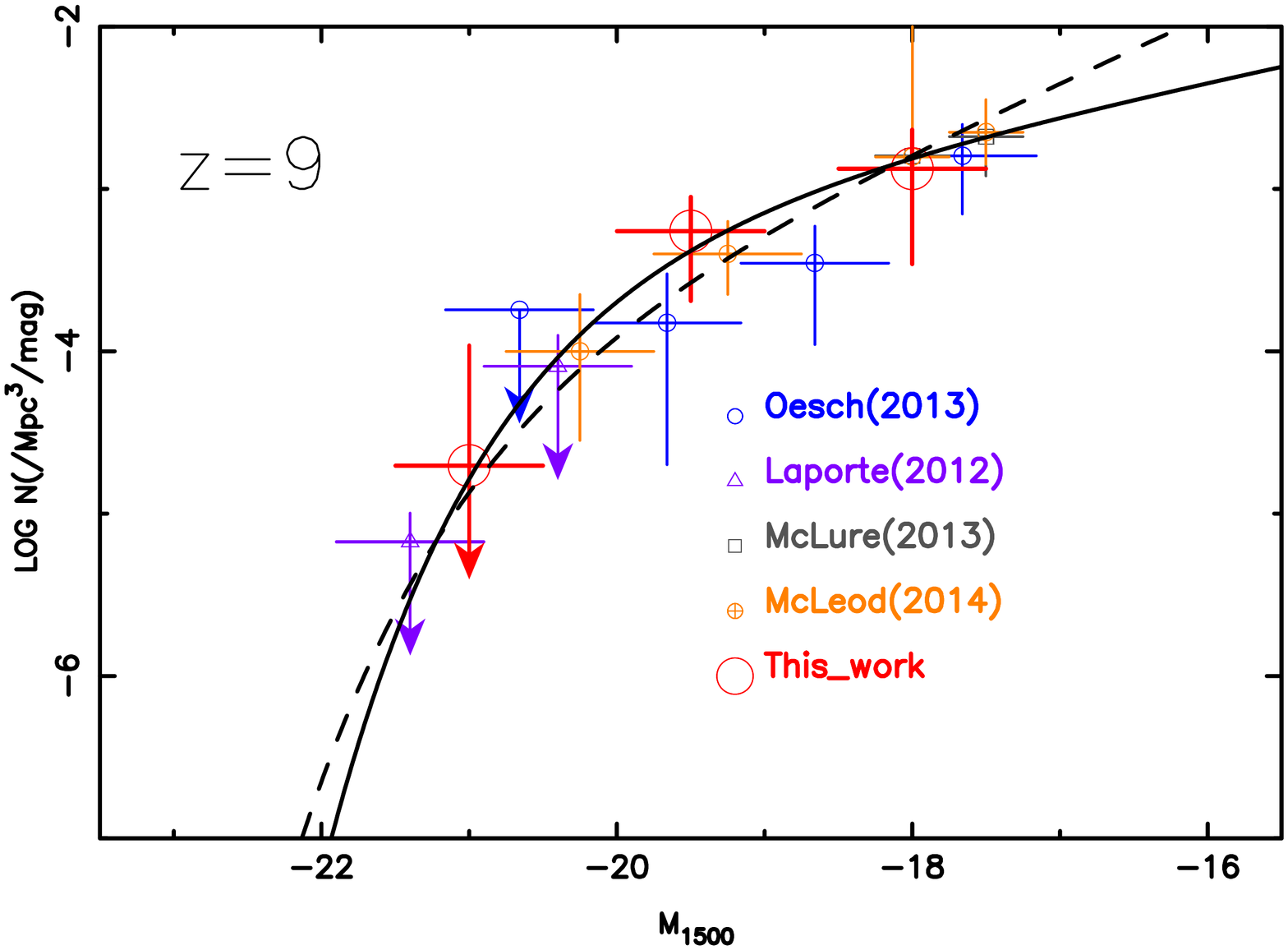}
           \includegraphics[width=6cm,angle=0]{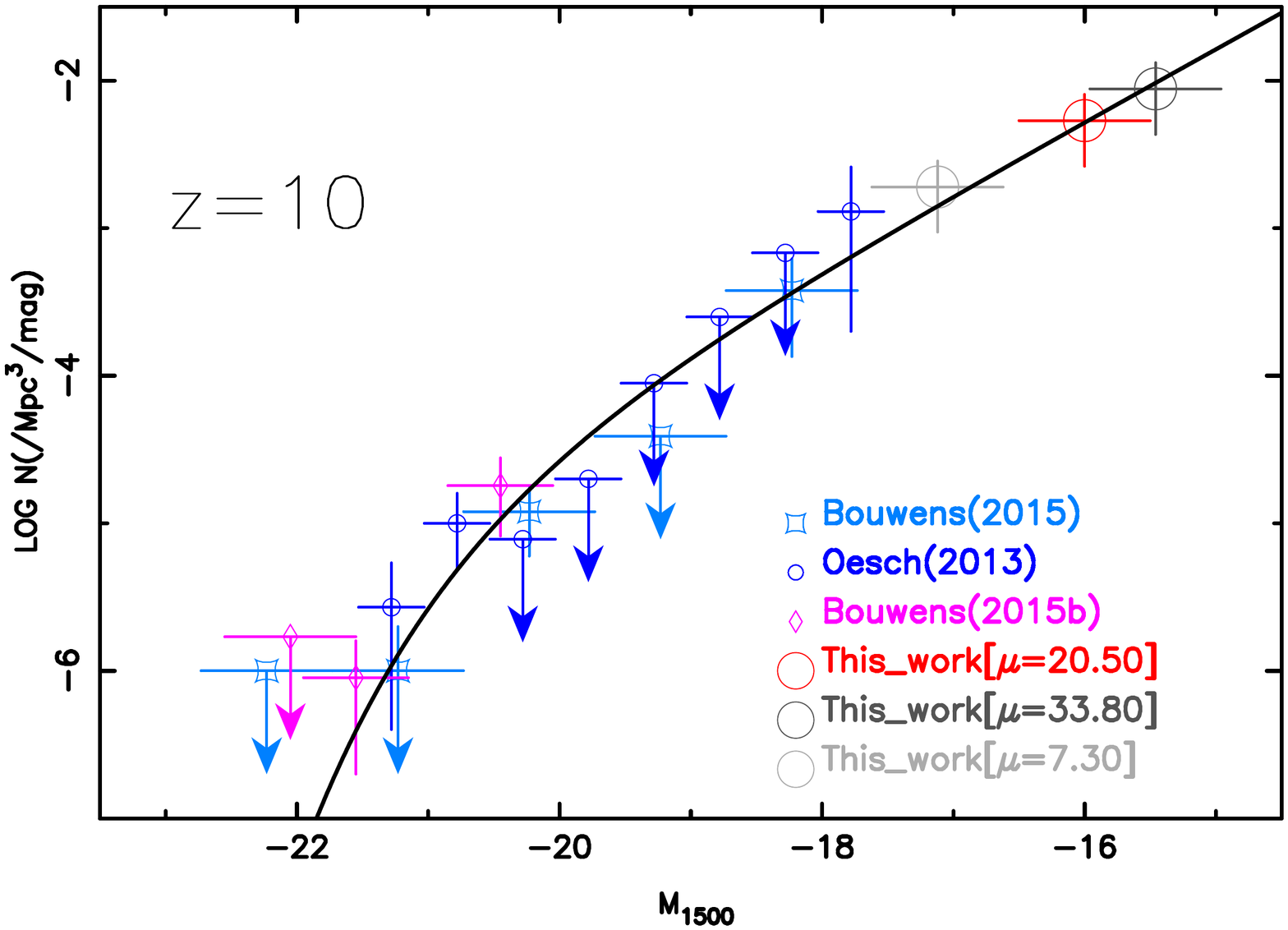}\\
      \caption{Shape of the luminosity functions computed
at $z\sim$7, 8, 9 and 10 using samples selected from the MACS0416 cluster and parallel
fields of the \HFF s. In each case, the black
line displays the best fit found using the densities computed in this
study and other densities at brightest luminosities published by other
groups (see below). \textit{Upper left} : LF at $z\sim$7 including
results from \citet{Schenker}, \citet{McLure}, \citet{bouwens4} and
\citet{Bowler}. The dashed line shows the parameterization computed by
\citet{bouwens4}. \textit{Upper right} : LF at $z\sim$8 including
results from \citet{Schenker}, \citet{oesch12},\citet{laporte2}
\citet{bouwens4},\citet{bradley} and \citet{atek}. The dashed line
shows the parameterization computed by \citet{bouwens4}.
\textit{Lower left} : LF at $z\sim$9 including results from
\citet{Bouwens6}, \citet{oesch12}, \citet{Laporte12} \citet{zheng} and
\citet{McLure}.  The dashed line shows the parameterization computed
by \citet{McLeod}.  \textit{Lower right} : LF at $z\sim$10 including
results from \citet{oesch14} and \citet{bouwens4}. The dashed line
shows the parameterization assuming the $\alpha$ parameter found by
\citet{bouwens4}.}
 \label{fig:LF} 
 \end{figure}

We adopted the Schechter parametrization \citep{schechter} to study
the evolution of the shape of the UV LF from $z\sim$7 to 10. The three
parameters were deduced using a $\chi^2$ minimization approach using
the densities estimated in this study combined with previous
results. Errors bars at $z\sim$7, 8 and 9 were deduced from the
1$\sigma$ confidence interval. Given the paucity of objects currently
known above $z\sim$10, we fixed the value of M$^{\star}$ for the
$z\sim$10 LF to that adopted by \citet{bouwens15}. Our parameterization
is consistent with what is expected from the UV LF evolution \citep{bouwens4}.
The shapes of the different
LFs are shown on Figure~\ref{fig:LF} and the values of the three Schechter
parameters are given in Table~\ref{tbl:LF}.

{\bf At $z=$10, our candidate is located in a highly magnified region leading to a large 1$\sigma$ error bar on the amplification. 
We took into account these uncertainties by computing the corresponding number densities over the amplification range covered by the
1$\sigma$  interval. Figure~\ref{fig:LF} shows that the 3 constraints we deduce follow the same shape. However to not bias the estimation
of the Schechter parameters, we only fitted the UV LF using the density computed assuming the mean amplification ($\mu\sim$20.50). 
We note that the best fit follows the shape described by both densities computed assuming different amplification values.}

\subsection{Comparison with Previous Studies}

\begin{figure}
   \centering
           \includegraphics[width=12cm,angle=-90]{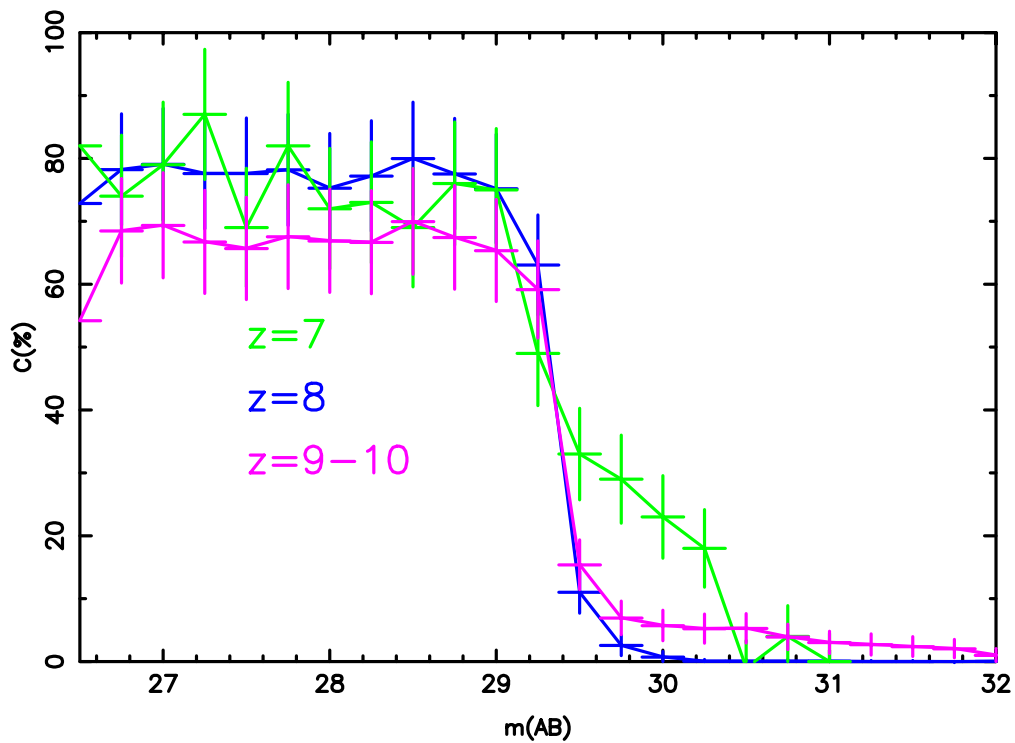}
      \caption{ Completeness levels of our samples as a function of magnitude in the filter tracing the $L_{1500}$ luminosity based on a mock catalog containing 680 000 objects with magnitude ranging from 22 to 31 with redshifts ranging from 6.5--10.5 and assuming a log-normal distribution for the size of each object (see text for details). Errors bars are deduced from the standard Poison uncertainties. }
\label{fig:completeness}
\end{figure}

The \HFF\ data for MACS0416 have been available since September 2014,
and two papers focusing on the brightest objects at $z\sim8$ and 9
have been published recently (\citealt{laporte2} and
\citealt{McLeod}). A key difference between our selection and the one
used in \citet{laporte2} is the redshift interval covered by the
color-criteria. The selection window we used to select the highest
redshift objects is well-defined over a redshift range between 6.5 and
10.5 (c.f. Figure~\ref{fig:completeness}) whereas Figure~2 of
\citet{laporte2} shows that their window is primarily devoted to the
selection of objects between 7.5 and 8.5. We note that two objects in
our sample were not selected in \citet{laporte2}: one of
these, 1859, fails to satisfy the magnitude cut they applied, which
only selects relatively bright candidates in the field.
\citealt{McLeod} selected five objects using a purely photometric
redshift approach, among them four are in common with the sample
described in this paper (491, 8428, 1213 and 1859 in our sample),
while one object is not (ID = HFF2C-9-5). Note that our candidate 4008
is not in the redshift range they targeted.

\section{CONCLUSIONS}

In order to study the properties of young galaxies in the era of
reionization, we characterize 22 galaxies between $z \sim 7$-10
discovered in the \HFF M0416 cluster ($z = 0.396$) and parallel fields.
Four of these galaxies were previously reported by \cite{laporte2} and
four by \citet{McLeod}. Our selection employed data from the
\HST\ optical, $J$ and $H$ bands, as well as deep HAWKI/VLT $K_s$ and
\ssp\/IRAC 3.6 and 4.5 \micron\ images. To find candidates, we carried
out careful and strict color selection. Simulations indicate that the completness in 
our sample ranges between 65\% and 85\% at redshifts between 7 and
10.0.

Photometric redshifts were calculated using a new version of the
Bayesian Photometric Redshifts code (BPZ2.0). We develop a new quality criteria based on
the ratio between peak early and late galaxy type probabilities
($\Re$), and chose candidates with peak-probability ratios $\Re
\geqslant 3$. Furthermore, to test the reliability of our redshift
estimations we used {\tt iSEDfit}, which produced very similar
results, with a mean difference of only $\Delta z=0.1\pm0.1$ ({\tt
  iSEDfit} minus BPZ2.0).

Magnifications are computed for the different candidates at the 
corresponding redshifts from an combination of six well selected 
models available in the MAST archive\footnotemark[20]. We looked for
counter images for objects near critical curves. In the case of 8958, 
we search for images at two positions predicted by our models. Although
there are some extremely faint red objects at these positions, no conclusive
identifications could be reached.

Based on an SED analysis with {\tt iSEDfit}, we computed stellar masses,
star formation rates and ages for each candidate.  Our $z>7$ candidates
are consistent with having mean stellar masses of
$log(M_*)8.40^{+0.55}_{-0.31}$~\Msun, SFRs of approximately
$1.6^{+0.5}_{-0.4} $~\Msun~yr$^{-1}$, and SFR-weighted ages of
$\lesssim 310^{+70}_{-140} $~Myr. 
These measurements agree with the results of \cite{schaerer10}.

Finally, based on candidates in M0416 cluster and parallel fields, we
computed LFs at $z = 7, 8, 9$ \& 10. 
Thanks to the depth of \HFF s, we are able to give the first {\bf direct estimates}
on the faint-end of the UV LF at $z\sim$10. {\bf The constraints given
in Table~\ref{tbl:LF} are clearly stronger than in previous studies.} This confirms
the crucial role of the \HFF s to detect object beyond the limit of
current telescopes. The Schechter parameters are relatively
well-constrained at $z\sim 7$ and 8 regarding the number of objects in
each samples. At higher redshifts, however, the full Frontier Fields
dataset will be required to increase the number of sources and 
reduce the cosmic variance effects.

\hskip 1.5in

\section{ACKNOWLEDGMENTS}

The work presented in this paper is based on observations made with
the NASA/ESA {\it Hubble Space Telescope}, and has been supported by
award AR-13279 from the Space Telescope Science Institute (STScI),
which is operated by the Association of Universities for Research in
Astronomy, Inc. under NASA contract NAS~5-26555.  It is also based on
data obtained with the {\it Spitzer Space Telescope}, which is
operated by the Jet Propulsion Laboratory, California Institute of
Technology under a contract with NASA.  We acknowledge support from
Basal-CATA PFB-06/2007 (LI, FEB, NL, SK, PT),
FP7-SPACE-2012-ASTRODEEP-312725 (XS), NSFC grants 11103017 and
11233002 (XS), NASA through Hubble Fellowship grant HST-HF2-51334.01-A
awarded by STScI (AZ), CONICYT-Chile grants FONDECYT 1141218 (FEB),
Gemini-CONICYT 32120003 (FEB, NL), and ``EMBIGGEN''
Anillo ACT1101 (FEB), the Ministry of Economy, Development, and
Tourism's Millennium Science Initiative through grant IC120009,
awarded to The Millennium Institute of Astrophysics, MAS (FEB),
Spanish consolider project CAD2010-00064 and AYA2012-39475-C02-01
(JMD), the Brazilian funding agency-FAPESP pos-doc fellowship - process number 2014/11806-9 (AM) and CONICYT-Chile grant FONDECYT 3140542 (PT).
LI would like to thank the European Southern Observatory and the University of Victoria, Canada,
for providing office space while this work was done.

\hskip 1.5in

%
%

\clearpage

%
%

\clearpage

\clearpage


\clearpage

\clearpage

\clearpage

\clearpage

\clearpage


%
%

%
{
\begin{deluxetable}{ccccc}
\tablecaption{Summary of Observations
\label{tbl:obs}}
\tablewidth{0pt}
\footnotesize
\tighttable
\tablehead{
\colhead{Telescope} &
\colhead{Band} &
\colhead{Exposure Time} &
\colhead{Limiting}
\\
\colhead{} &
\colhead{} &
\colhead{(sec)} &
\colhead{Magnitude ($5\sigma$)}
}
\startdata 
&&M0416&\\
\hline
\HST  &  F435W  &     54512  &    29.2\\ 
\HST  &  F606W  &     35450  &    29.2\\ 
\HST  &  F814W  &   131897  &    29.3\\ 
\HST  &  F105W   &    72567  &    29.0\\ 
\HST  &  F125W   &    39302  &    28.7\\
\HST  &  F140W  &     29870  &    28.6\\ 
\HST  &  F160W  &     74787  &    28.8\\ 
VLT/HAWKI& $K_s$    &   97440	  &    26.5\\ 
Spitzer & IRAC1 &  340712 & 25.3 \\
Spitzer & IRAC2 &  340712 & 25.3 \\
\hline
&&Parallel field &\\ 
\hline
\HST  &  F435W  &     45747  &    29.0\\ 
\HST  &  F606W  &     25035  &    28.9\\ 
\HST  &  F814W  &     95406  &    29.2\\
\HST  &  F105W   &    79912  &    29.1\\ 
\HST  &  F125W   &    34248  &    28.5\\ 
\HST  &  F140W   &    34248  &    28.6\\ 
\HST  &  F160W  &     79912  &    28.8\\ 
VLT/HAWKI& $K_s$    &   97440  &    26.5\\
Spitzer & IRAC1 &  340712 & 25.3 \\
Spitzer & IRAC2 &  340712 & 25.3 \\
\enddata
\end{deluxetable}
}

\clearpage

{
\begin{deluxetable}{lcc}
\tablecaption{IRAC photometry for Selected Candidates\tablenotemark{a}
\label{tbl:IRAC}}
\tablewidth{0pt}
\tiny 
\tighttable
\tablehead{
\colhead{Name} &
\colhead{IRAC 1} &
\colhead{IRAC 2}
\\
\colhead{} &
\colhead{} &
\colhead{} 
}
\startdata
\hline
&M0416&\\
\hline
         8958  &         $>  25.31$  				&     $>  25.75$\\
         1859  &         $>  26.67$ 				&      $>  27.11$\\
         8364\tablenotemark{b}  &         $26.59 \pm 0.52$  	&    $25.79 \pm 0.23$\\
          491  &          $>  27.07$   			&   $24.72 \pm 0.20$\\
         1213 &          $>  26.67$   			&    $>  26.76$\\
         8428\tablenotemark{b}  &          $26.73 \pm 0.60$  	&    $25.93 \pm 0.26$\\
         4008 &          $>  25.32$  				&    $>  25.54$\\
\hline
&Parallel Field&\\
\hline
         3687 &   $>  27.54$  &  $>  27.76$\\
         4177 &   $>  26.41$  &  $>  26.59$\\
         3076 &   $>  26.92$  &  $>  27.69$\\
         1301 &   $>  27.22$  &  $>  27.45$\\
         3814 &   $>  27.52$  &  $>  27.10$\\
         1241 &   $>  27.22$  &  $>  27.45$\\
         5296\tablenotemark{c} &   $23.22 \pm 0.73$ &  $22.66 \pm  0.02$\\
         3790 &   $>  26.64$  &  $>  26.70$\\
         4125 &   $26.65 \pm  0.30$  &  $26.58 \pm 0.25$\\
         6999 &   $>  27.30$  &  $>  27.40$\\
         7361 &   $>  27.32$  &  $>  27.27$\\
         1331 &   $>  27.93$  &  $>  27.78$\\
         1386 &   $>  26.72$  &  $>  27.01$\\
         1513 &   $26.96 \pm  0.44$ &   $26.62 \pm 0.26$\\
          146  &   $>  27.26$  &  $>  27.85$\\
\hline
\enddata 
\tablenotetext{a}{The symbol "$>$" represents 1 $\sigma$ upper limits on no-detection.} 
\tablenotetext{b}{8364 and 8428 are within 2 pixels from each other in IRAC images. We estimated their IRAC flux based on the  F160W band flux ratio.} 
\tablenotetext{c}{This object is contaminated by a nearby spiral galaxy. The photometry is largely uncertain.}
\end{deluxetable}
}

\clearpage

%
{
\begin{deluxetable}{lcclclclclc}
\tablecaption{Photometric Redshift estimations 
\label{tbl:photoz}}
\tablewidth{0pt}
\tiny 
\tighttable
\tablehead{
\colhead{ID} &
\colhead{$z_{peak}^{late}$} &
\colhead{$z_{min}^{late}$} &
\colhead{$z_{max}^{late}$} &
\colhead{$z_{peak}^{early}$} &
\colhead{$z_{min}^{early}$} &
\colhead{$z_{max}^{early}$} &
\colhead{$\Re$} 
}
\startdata
\hline
M0416&&&&&&&\\
\hline
  8958 & 10.112 & 9.840  & 10.632 & 2.476 & 2.262 & 10.03 & 4.6    \\     
  1859 & 9.354 &  9.074 & 9.54  & 1.992 & 1.839 & 2.189 & 3.8   \\      
  8364 & 9.234 & 9.049 & 9.294 & --- & --- & --- & -1.0   $^{*}$  \\  
   491  & 8.478 & 8.422 & 8.540 & --- & --- & --- & -1.0 $^{*}$   \\     
  8428 & 8.353 &  8.253 & 8.392 & --- & --- & --- &  -1.0 $^{*}$  \\      
  1213 & 8.311  &  8.138 & 8.392 & 1.483 & 1.428 & 8.052 & 2.2e+6   \\ 
  4008 & 7.734 & 7.561 & 7.832 & 1.118 & 0.999 & 1.270  & 2.8e+4  \\    
\hline
Parallel&&&&&&&\\
\hline
 3687 & 9.354 &  8.804 & 9.656 & 2.027 & 2.048 & 9.105 & 4.7   \\      
 4177 & 9.354 &  8.995 & 9.439 & 1.992 & 1.810  & 2.095 & 9.3    \\      
 3076 & 9.002 &  8.837 & 9.249 & 8.807 & 8.697 & 9.03 & 567.5   \\    
 5296 & 8.360 &  8.356 & 8.371 & --- & --- & --- & -1.0  $^{*}$ \\      
 1301 & 8.334 &  8.098 & 8.444 & 1.512 & 1.354 & 1.766 & 783.7  \\     
 3814 & 8.126 & 7.816 & 8.342 & 1.570 & 1.372 & 1.815 & 5.64   \\      
  1241 & 7.939 & 7.738 & 8.133 & 1.399 & 1.301 & 1.519 & 5.8e+4  \\    
 3790 & 7.793 & 7.674 & 7.879 & --- &  --- & --- &  -1.0 $^{*}$  \\      
  4125 & 7.711 & 7.413 & 7.808 & 1.118 & 0.991 & 1.289 & 9.7   \\       
 6999 & 7.634 &  7.489 &  7.695 & ---  &  --- &  --- &  -1.0 $^{*}$ \\       
 7361 & 7.513 &  7.172 & 7.688 & 0.875 & 1.251 & 1.466 & 46.1  \\      
  1331 & 7.248 & 7.147 &  7.374 & 0.289 & 0.161 & 0.289 & 8.9      \\   
 1386 & 7.243  & 7.096 & 7.357 & 0.289 &  0.162 & 0.348 & 35.4    \\    
 146  & 7.151  & 6.948 & 7.326 & --- &  --- &  --- &  -1.0 $^{*}$   \\     
 1513 & 7.000 & 6.813 & 7.178 & 1.265 & 1.150  & 1.291 & 1.6e+7  \\      
\hline
\enddata 
\tablenotetext{1}{The table list the photometric redshift estimations for the 22 photometrically selected candidates, where $z_{peak}^{late}$ and $z_{peak}^{early}$ represent the peak-values for the late- and early-type spectral-solutions (respectively), $z_{min}$ and $z_{max}$ correspond to a $1\sigma$ confident interval and $\Re$ the peak-probability ratio among solutions. }
\tablenotetext{2}{Asterisked ratios (*) correspond to detections without low-redshift solutions.}
\end{deluxetable}
}
\clearpage

{
\begin{deluxetable}{lcccccccccc}
\tabletypesize{\tiny}
\tabletypesize{\scriptsize}
\setlength{\tabcolsep}{0.05in}
\rotate
\tablecaption{$z > 7$ Candidates and Their Properties: M0416 Field.
\label{tbl:z7}}
\tablewidth{0pt}
\tiny 
\tighttable
\tablehead{
\colhead{Name} &
\colhead{Photometric} &
\colhead{RA} &
\colhead{Dec} &
\colhead{F160W} &
\colhead{F140W} &
\colhead{F125W} &
\colhead{F105W} &
\colhead{F814W} &
\colhead{$K_s$\tablenotemark{b}} &
\colhead{$\mu$\tablenotemark{c}} \\
\colhead{} &
\colhead{Redshift} &
\colhead{(J2000)} &
\colhead{(J2000)} &
\colhead{} &
\colhead{} &
\colhead{} &
\colhead{} &
\colhead{} &
\colhead{} &
\colhead{} 
}
\startdata
8958	& $10.112$ & $64.025406$ & $-24.082251$ & $28.50 \pm 0.11$ & $28.99 \pm 0.17$ & $ >30.40$ & $31.86 \pm 1.92$ & $ >30.70$ &  $ >27.03$ & $20.50 \pm 13.30$\\

1859	& $ 9.354$ & $64.043114$ & $-24.057899$ & $27.72 \pm 0.07$ & $27.93 \pm 0.09$ & $28.63 \pm 0.18$ & $30.53 \pm 0.79$ & $30.14 \pm 0.56$ &  $27.21 \pm	0.42$ & $4.41 \pm 2.44$\\

8364	& $ 9.234$ & $64.048058$ & $-24.081429$ & $26.45 \pm 0.03$ & $26.62 \pm 0.04$ & $27.07 \pm 0.06$ & $29.99 \pm 0.56$ & $ >30.70$ &  $26.52 \pm	0.22$ & $1.80 \pm 0.47$\\

491 	& $ 8.478$ & $64.039169$ & $-24.093187$ & $25.92 \pm 0.02$ & $25.99 \pm 0.03$ & $26.41 \pm 0.04$ & $27.90 \pm 0.10$ & $28.96 \pm 0.34$ & $26.40 \pm	0.20$ & $1.71 \pm 0.47$\\

8428	& $ 8.353$ & $64.047981$ & $-24.081665$ & $26.59 \pm 0.03$ & $26.59 \pm 0.03$ & $26.73 \pm 0.04$ & $28.19 \pm 0.10$ & $ >30.70$ & $ >27.54$ & $1.79 \pm 0.46$\\

1213	& $ 8.311$ & $64.037582$ & $-24.088104$ & $28.12 \pm 0.08$ & $27.80 \pm 0.07$ & $27.95 \pm 0.08$ & $29.51 \pm 0.25$ & $33.03 \pm 7.60$ & $>27.46$ & $2.36 \pm 1.01$\\

4008	& $ 7.734$ & $64.060333$ & $-24.064960$ & $27.85 \pm 0.08$ & $27.86 \pm 0.09$ & $27.53 \pm 0.07$ & $28.48 \pm 0.11$ & $31.47 \pm 1.57$ & $ >27.54$ & $2.17 \pm 0.31$\\
\enddata
\tablenotetext{a}{Magnitudes are isophotal,
scaled by an aperture correction term derived in the \H\ band. The errors and 
limiting magnitudes are $1\sigma$. Photometric redshifts have been derived using BPZ2.0, and the quoted
uncertainties indicate the 68\% confidence interval.}
\tablenotetext{b}{Photometry within a 0.4 \arcsec radius aperture.}
\tablenotetext{c}{See \S \ref{sec:lens} for magnification factors and errors.}
\end{deluxetable}
}

\clearpage

{
\begin{deluxetable}{lcccccccccc}
\tabletypesize{\tiny}
\tabletypesize{\scriptsize}
\setlength{\tabcolsep}{0.05in}
\rotate
\tablecaption{$z > 7$ Candidates and Their Properties: M0416 Parallel Field.
\label{tbl:z7p}}
\tablewidth{0pt}
\tiny 
\tighttable
\tablehead{
\colhead{Name} &
\colhead{Photometric} &
\colhead{RA} &
\colhead{Dec} &
\colhead{F160W} &
\colhead{F140W} &
\colhead{F125W} &
\colhead{F105W} &
\colhead{F814W}&
\colhead{$K_s$\tablenotemark{b}} \\
\colhead{} &
\colhead{Redshift} &
\colhead{(J2000)} &
\colhead{(J2000)} &
\colhead{} &
\colhead{} &
\colhead{} &
\colhead{} &
\colhead{} &
\colhead{} 
}
\startdata

3687  & $ 9.354$ & $64.156128$ & $-24.111143$ & $28.84 \pm	 0.14$ & $29.19 \pm 0.19$ & $29.63 \pm 0.32$ & $ >31.30$ & $32.37 \pm 4.17$ & $27.78 \pm 0.71$\\	

4177  & $ 9.354$ & $64.149864$ & $-24.113352$ & $27.56 \pm	 0.06$ & $27.61 \pm 0.06$ & $28.44 \pm 0.15$ & $29.65 \pm 0.38$ & $31.81 \pm 3.82$ & $ >28.35$\\	

3076  & $ 9.002$ & $64.151779$ & $-24.108454$ & $27.84 \pm	 0.07$ & $27.95 \pm 0.08$ & $28.39 \pm 0.14$ & $ >31.30$ & $ >31.40$ & $ >28.08$\\	

5296  & $ 8.360$ & $64.131889$ & $-24.119350$ & $24.18 \pm	 0.01$ & $24.49 \pm 0.01$ & $24.99 \pm 0.02$ & $26.19 \pm 0.04$ & $ >31.40$ & $24.03 \pm 0.03$\\	

1301  & $ 8.334$ & $64.126785$ & $-24.100315$ & $28.23 \pm	 0.08$ & $28.10 \pm 0.07$ & $28.30 \pm 0.09$ & $29.69 \pm 0.27$ & $30.21 \pm 0.67$ & $ >28.70$\\	

3814  & $ 8.126$ & $64.147903$ & $-24.111727$ & $28.12 \pm	 0.08$ & $28.19 \pm 0.09$ & $28.34 \pm 0.12$ & $29.56 \pm 0.29$ & $31.29 \pm 1.93$ & $26.74 \pm 0.27$\\	

1241 & $7.939$ & $64.126862$ & $-24.100067$ & $28.15 \pm 0.07$ & $27.89 \pm 0.06$ & 	$28.10 \pm	0.08$ & 	$29.03 \pm 0.15$ & $	 > 31.40$ & $ >28.03$\\	

3790  & $ 7.793$ & $64.136993$ & $-24.111664$ & $26.40 \pm	 0.03$ & $26.54 \pm 0.03$ & $26.69 \pm 0.04$ & $27.61 \pm 0.08$ & $ >31.40$ & $ >27.74$\\	

4125 & $7.711$	 & $64.120026$ & $-24.113226$ & $27.82 \pm 0.06$ & 	$27.89 \pm	0.07$ & $27.80 \pm 0.07$ &  $28.60 \pm	 0.12$ & $29.85 \pm 0.47$ & $ >28.19$\\	

6999  & $ 7.634$ & $64.121368$ & $-24.125864$ & $27.33 \pm	 0.04$ & $27.40 \pm 0.05$ & $27.22 \pm 0.04$ & $27.94 \pm 0.07$ & $ >31.40$ & $ >28.30$\\	

7361 & $ 7.513$ & $64.137886$ & $-24.127241$ & $28.22 \pm	 0.08$ & $28.07 \pm 0.07$ & $28.34 \pm 0.11$ & $28.89 \pm 0.14$ & $ >31.40$ & $ >28.50$\\	

1331 & $7.248$	 & $64.144447$ & $-24.100300$ & $27.37 \pm 0.05$ & 	$27.35 \pm	 0.05$ & $27.24 \pm 0.05$ & 	$27.68 \pm	 0.06$ & $29.38 \pm	0.39$ & $ >28.06$\\

1386 & $ 7.243$ & $64.149895$ & $-24.100410$ & $27.82 \pm	 0.06$ & $27.62 \pm 0.05$ & $27.66 \pm 0.06$ & $28.04 \pm 0.06$ & $31.21 \pm 1.51$ & $ >28.32$\\	

  146 & $7.151$ & $64.130882$ & $-24.092710$ & $27.58 \pm 0.06$ & $27.50 \pm 0.05$ & $27.72 \pm  0.07$ & $27.99 \pm 0.07$ & $ >	31.40$ & $ >28.19$\\	

1513 & $ 7.000$ & $64.142899$ & $-24.101274$ & $28.14 \pm	 0.08$ & $28.12 \pm 0.08$ & $28.02 \pm 0.08$ & $28.30 \pm 0.09$ & $ >31.40$ & $ >28.58$\\	
\hline
\enddata
\tablenotetext{a}{Magnitudes are isophotal,
scaled by an aperture correction term derived in the \H\ band. The errors and 
limiting magnitudes are $1\sigma$. Photometric redshifts have been derived using BPZ2.0, and the quoted
uncertainties indicate the 68\% confidence interval.}
\tablenotetext{b}{Photometry within a 0.4 \arcsec radius aperture.}
\end{deluxetable}
}	

\clearpage

%

\begin{table*}[!ht]
\caption{Physical properties of the candidates inferred from their SED.}
\label{tbl:pp}
{\centering
\begin{tabular}{lcccc}
\hline\hline
\noalign{\smallskip}
Galaxy ID & Redshift & $\rm log M_*$   	& log SFR  	& log AGE				\\
          &          & $\rm [M_{\odot}]$  & $\rm [M_{\odot} yr^{-1}]$  &$\rm [Gyr]$      \\
\hline
\noalign{\medskip}
\hline
    M0416&         &                                  &                                  &                              \\ 
\hline
    8958 &   10.112 &   8.92$\rm ^{+0.50}_{-0.29   } $ &  -0.13$\rm ^{+0.13}_{-0.60   } $ &   0.33$\rm ^{+0.09}_{-0.13} $\\ 
    1859 &    9.354 &   8.52$\rm ^{+0.28}_{-0.43   } $ &   0.37$\rm ^{+0.09}_{-0.08   } $ &   0.25$\rm ^{+0.17}_{-0.17} $ \\
    8364 &    9.234 &   9.15$\rm ^{+0.28}_{-0.37   } $ &   0.86$\rm ^{+0.07}_{-0.09   } $ &   0.32$\rm ^{+0.14}_{-0.17} $ \\
     491 &    8.478 &   9.49$\rm ^{+0.17}_{-0.22   } $ &   0.95$\rm ^{+0.07}_{-0.13   } $ &   0.40$\rm ^{+0.13}_{-0.16} $ \\
    8428 &    8.353 &   9.12$\rm ^{+0.26}_{-0.31   } $ &   0.73$\rm ^{+0.06}_{-0.10   } $ &   0.38$\rm ^{+0.15}_{-0.18} $ \\
    1213 &    8.311 &   8.40$\rm ^{+0.30}_{-0.42   } $ &   0.22$\rm ^{+0.08}_{-0.08   } $ &   0.29$\rm ^{+0.20}_{-0.19} $ \\
    4008 &    7.734 &   8.44$\rm ^{+0.32}_{-0.44   } $ &   0.26$\rm ^{+0.09}_{-0.08   } $ &   0.30$\rm ^{+0.23}_{-0.21} $ \\
\hline
    Parallel &     &                                  &                                  &                              \\ 
\hline
    3687 &    9.354 &   8.05$\rm ^{+0.33}_{-0.46   } $ &  -0.10$\rm ^{+0.10}_{-0.10   } $ &   0.25$\rm ^{+0.17}_{-0.17} $ \\
    4177 &    9.354 &   8.59$\rm ^{+0.31}_{-0.42   } $ &   0.46$\rm ^{+0.09}_{-0.08   } $ &   0.25$\rm ^{+0.17}_{-0.16} $ \\
    3076 &    9.002 &   8.34$\rm ^{+0.22}_{-0.34   } $ &   0.29$\rm ^{+0.08}_{-0.07   } $ &   0.22$\rm ^{+0.16}_{-0.13} $ \\
    5296 &    8.360 &  11.02$\rm ^{+0.04}_{-0.04   } $ &   0.99$\rm ^{+0.18}_{-0.40   } $ &   0.34$\rm ^{+0.15}_{-0.11} $ \\
{\bf5296}&    6.901 &  11.04$\rm ^{+0.01}_{-0.01   } $ &  -3.83$\rm ^{+0.24}_{-0.24   } $ &   0.27$\rm ^{+0.02}_{-0.02} $ \\
    1301 &    8.334 &   8.31$\rm ^{+0.29}_{-0.42   } $ &   0.13$\rm ^{+0.08}_{-0.08   } $ &   0.28$\rm ^{+0.20}_{-0.18} $ \\
    3814 &    8.126 &   8.38$\rm ^{+0.30}_{-0.40   } $ &   0.08$\rm ^{+0.08}_{-0.10   } $ &   0.33$\rm ^{+0.19}_{-0.20} $ \\
    1241 &    7.939 &   8.37$\rm ^{+0.27}_{-0.39   } $ &   0.14$\rm ^{+0.08}_{-0.08   } $ &   0.31$\rm ^{+0.20}_{-0.19} $ \\
    3790 &    7.793 &   8.67$\rm ^{+0.09}_{-0.14   } $ &   0.77$\rm ^{+0.04}_{-0.06   } $ &   0.16$\rm ^{+0.07}_{-0.06} $ \\
    4125 &    7.711 &   8.58$\rm ^{+0.38}_{-0.41   } $ &   0.20$\rm ^{+0.08}_{-0.12   } $ &   0.39$\rm ^{+0.19}_{-0.23} $ \\
    6999 &    7.634 &   8.29$\rm ^{+0.19}_{-0.31   } $ &   0.47$\rm ^{+0.08}_{-0.06   } $ &   0.14$\rm ^{+0.12}_{-0.09} $ \\
    7361 &    7.513 &   8.36$\rm ^{+0.29}_{-0.40   } $ &   0.04$\rm ^{+0.08}_{-0.09   } $ &   0.37$\rm ^{+0.21}_{-0.22} $ \\
    1331 &    7.248 &   7.52$\rm ^{+0.10}_{-0.06   } $ &   0.83$\rm ^{+0.19}_{-0.13   } $ &   0.01$\rm ^{+0.01}_{-0.00} $ \\
    1386 &    7.243 &   8.49$\rm ^{+0.27}_{-0.38   } $ &   0.22$\rm ^{+0.08}_{-0.08   } $ &   0.35$\rm ^{+0.23}_{-0.21} $ \\
     146 &    7.151 &   8.08$\rm ^{+0.16}_{-0.27   } $ &   0.33$\rm ^{+0.07}_{-0.06   } $ &   0.12$\rm ^{+0.09}_{-0.07} $ \\
    1513 &    7.000 &   8.47$\rm ^{+0.38}_{-0.43   } $ &   0.03$\rm ^{+0.09}_{-0.11   } $ &   0.44$\rm ^{+0.22}_{-0.26} $ \\
\noalign{\smallskip}
\hline\hline                 
\noalign{\smallskip}
\end{tabular}
}
\\
The following quantities are reported in each column:
 col. 1, object name;
 col. 2, object redshift;
 col. 3, logarithm of the stellar mass inferred by using the FSPS-v2.4 miles;
 col. 4, star formation rate ;
 col. 5, galaxy age.
 Values corrected by their magnification factor. Errors are shown at $1\sigma$ confidence level.
\end{table*}

\clearpage

\begin{table}
\caption{Number densities at $z\sim$7, 8,  9 and 10.}             
\label{tbl:densities}      
\begin{tabular}{c | cc  l}        
\hline\hline                 
Redshift Range & M$_{1500}$	& $\Phi(M_{1500})$	\\	
			& 			& [/Mpc$^3$/mag)	\\
 \hline                        
 {$z\sim$7}	& -20.75 $\pm$ 0.500 & (1.64$\pm$1.37)
$\times$10$^{-04}$ \\
							& -19.75 $\pm$ 0.500 & (3.72$\pm$2.32) $\times$10$^{-04}$ \\
							& -18.75 $\pm$ 0.500 & (1.03$\pm$0.44) $\times$10$^{-03}$ \\
							& -17.50 $\pm$ 0.500 & (7.10$\pm$1.95) $\times$10$^{-03}$ \\ \hline
 {$z\sim$8}	& -21.00 $\pm$ 0.500 & (5.34$\pm$5.34)
$\times$10$^{-05}$ \\
							& -20.00 $\pm$ 0.500 & (2.70$\pm$2.06) $\times$10$^{-04}$ \\
							& -19.25 $\pm$ 0.250 & (3.15$\pm$2.27) $\times$10$^{-04}$ \\
							& -18.25 $\pm$ 0.250 & (1.97$\pm$0.78) $\times$10$^{-03}$ \\
							& -17.50 $\pm$ 0.500 & (2.84$\pm$1.09) $\times$10$^{-03}$ \\ \hline
 {$z\sim$9}	& -21.00 $\pm$ 0.500 & (1.97$\pm$8.90)
$\times$10$^{-05}$ \\
							& -19.50 $\pm$ 0.500 & (5.49$\pm$3.45) $\times$10$^{-04}$ \\
							& -18.00 $\pm$ 0.500 & (1.33$\pm$0.99) $\times$10$^{-03}$ \\ \hline
 {$z\sim$10}	&  -16.00 $\pm$ 0.500 & (5.35$\pm$2.72)
$\times$10$^{-03}$ \\ \hline
\end{tabular}
\tablenotetext{}{Number densities were computed taking into account the redshift
probability distribution for each candidate. Error bars included Poisson
uncertainties and Cosmic Variance computed from
\citet{cv} and the effective surface of our survey.  }
\end{table}

\clearpage

\begin{table}
\caption{Schechter parameterization at $z\sim$7, 8,  9 and 10.}           
\label{tbl:LF}      
\begin{tabular}{c | ccc l}        
\hline\hline                 
Redshift 	&	M$^{\star}$					& 	$\alpha$				&	$\Phi^{\star}$ 	\\
		&	[AB]							&						& [/mag/Mpc$^{-3}$]	\\ \hline
z$\sim$7	&	-20.49
$^{+0.15}_{-0.11}$			&	-2.00$^{+0.13}_{-0.12}$	&	(0.41$^{+0.13}_{-0.14}$)$\times$10$^{-3}$
\\
z$\sim$8	&	-20.11
$^{+0.59}_{-0.25}$			&	-1.76$^{+0.35}_{-0.54}$	&	(0.56$\pm$0.39)$\times$10$^{-3}$
\\
z$\sim$9	&	-19.66
$^{+0.77}_{-0.44}$			&	-1.48$^{+0.32}_{-0.50}$	&	(1.00$^{+0.93}_{-0.77}$)$\times$10$^{-3}$
\\
z$\sim$10	&	-20.28 (\textit{fixed})	&	-2.23$^{+0.24}_{-0.05}$	&	(4.50$^{+1.70}_{-1.9}$)$\times$10$^{-5}$\\
\hline
\end{tabular}
\tablenotetext{}{Schechter parameters deduced from this study using a $\chi^2$
minimization method. Error bars are given by the 1$\sigma$ confidence
interval at $z\sim$7, 8, 9, and10. }
\end{table}

\clearpage

%

\end{document}